\documentclass[useAMS,usenatbib]{mn2e}

\usepackage{graphicx}
\usepackage{amssymb}
\usepackage{subfig}
\usepackage{tabularx}
\usepackage{threeparttable}
\newcommand{\eqb}{\begin{eqnarray}}
\newcommand{\eqe}{\end{eqnarray}}
\newcommand{\sth}{\sigma_{\rm T}}
\newcommand{\tth}{\tau_{\rm T}}
\newcommand{\tg}{t_{\gamma}}
\newcommand{\tcr}{t_{\rm cr}}
\newcommand{\dt}{\delta t}
\newcommand{\Lk}{L_{\rm k}}
\newcommand{\Lpinj}{L_{\rm p}^{\rm inj}}
\newcommand{\lp}{\ell_{\rm p}^{\rm inj}}
\newcommand{\lpcr}{\ell_{\rm p,cr}}
\newcommand{\lph}{\ell_{\gamma}}
\newcommand{\lcr}{\ell_{\gamma, \rm cr}}
\newcommand{\rb}{r_{\rm b}}
\newcommand{\eb}{\epsilon_{\rm B}}
\newcommand{\ep}{\epsilon_{\rm p}}
\newcommand{\Bcr}{B_{\rm cr}}
\newcommand{\emx}{\epsilon_{\max}}
\newcommand{\gmx}{\gamma_{\max}}
\newcommand{\gsat}{\gamma_{\rm sat}}
\newcommand{\gh}{\gamma_{\rm H}}
\newcommand{\xmn}{x_{\min}}
\newcommand{\xmx}{x_{\max}}
\newcommand{\xstar}{x_{\star}}
\newcommand{\gc}{\gamma_{\rm c}}
\newcommand{\mpr}{m_{\rm p}}
\newcommand{\mel}{m_{\rm e}}
\newcommand{\nph}{n_{\gamma}}
\newcommand{\fl}{f_{\ell}}

\title[Hadronic supercriticality \& GRB emission]
{Hadronic supercriticality  as a trigger for GRB emission}
\author[M.P., S.D., A.M., D.G.]{M. Petropoulou$^{1,2}$ \thanks{E-mail:mpetropo@purdue.edu (MP)}, S. Dimitrakoudis$^{3}$\thanks{E-mail: sdimis@noa.gr (SD)},
A. Mastichiadis$^{4}$ \thanks{E-mail: amastich@phys.uoa.gr (AM)}, D. Giannios$^{1}$\thanks{E-mail:  dgiannio@purdue.edu  (DG)}\\
$^{1}$Department of Physics and Astronomy, Purdue University, 525 Northwestern Avenue, West Lafayette, IN, 47907, USA \\
$^2$NASA Einstein Postdoctoral Fellow \\
$^{3}$Institute for Astronomy, Astrophysics, Space Applications \& Remote Sensing, National Observatory of Athens, 15 236 Penteli, Greece\\
$^{4}$Department of Physics, University of Athens, Panepistimiopolis, GR 15783 Zografos, Greece}

\begin{document}
\date{Received.../Accepted...}

\pagerange{\pageref{firstpage}--\pageref{lastpage}} \pubyear{2014}

\maketitle

\label{firstpage}
\begin{abstract}
We explore a one-zone hadronic model that may be able to reproduce $\gamma$-ray burst (GRB) prompt 
emission with a minimum of free parameters. Assuming only that GRBs are efficient high-energy proton 
accelerators and without the presence of an {\sl ab initio} photon field, 
we investigate the conditions under which the system becomes supercritical, i.e. there is a fast, non-linear transfer
of energy from protons to secondary particles initiated by the spontaneous quenching of proton-produced $\gamma$-rays. 
We first show analytically that the transition to supercriticality occurs whenever the proton injection compactness exceeds a critical value,
which favours high proton injection luminosities and a wide range of bulk Lorentz factors. 
The properties of supercriticality are then studied with a time-dependent numerical code that solves concurrently the coupled  
equations of proton, photon, electron, neutron and 
neutrino distributions. For conditions that drive the system deep into the supercriticality we find that the photon spectra 
obtain a Band-like shape due to Comptonization 
by cooled pairs and that the energy transfer efficiency from protons to $\gamma$-rays  
and neutrinos is high reaching $\sim 0.3$. Although some questions 
concerning its full adaptability to the GRB prompt emission remain open, 
supercriticality is found to be a promising process in that regard.

\end{abstract}

\section{Introduction}
\label{intro}
The prompt emission of gamma-ray bursts (GRBs) is  observed in the $10$~keV-$1$~MeV energy band \citep{preece00, fermiGBM14}, 
 with fluences that generally range between $10^{-4}$~erg~cm$^{-2}$ and $10^{-7}$~erg~cm$^{-2}$, where the lower limit does not
necessarily reflect an intrinsic property of GRBs but depends on the sensitivity of the detectors. 
 The fluence distribution of bursts  detected by the {\sl Fermi}/GBM peaks at $\sim 10^{-5}$~erg cm$^{-2}$ \citep{fermiGBM14}, which is considered as 
the typical GRB fluence.
GRB light curves are highly variable and have a complex structure: they consist 
of several pulses (10-100)  each of them having typical width $10$ms-$1$s \citep{norris96, nakarpiran02}, thus making the 
total duration of the burst longer, e.g. $10-100$~s. The GRB spectra can, in most cases, be described by 
a smoothly connected broken power-law \citep{band93, band09} with values of the break energy
clustered around $\sim 0.2-0.5$~MeV in the observer's frame, while
even higher peak energies ($>10$~MeV) have been detected (e.g. \citealt{goldstein12}).
The typical photon indices of the spectrum below and above the peak are $\alpha\sim -1$ and $\beta \sim -2.2$, respectively \citep{preece00, goldstein12}.

The origin of the GRB emission is still an open issue, although various models have been proposed during the past decades trying to address all or most
of the above properties. Borrowing the terminology from the field of blazar modelling (see  e.g. \citealt{boettcher07, boettcher10}, for reviews)
GRB emission models can be divided  in two categories, namely leptonic and hadronic, according to the species of the radiating particles.

The former try to attribute the gamma-ray emission by employing radiation processes of relativistic electrons (and/or positrons). 
The classical scenario of the prompt emission, which belongs to the first category, is the optically thin synchrotron model  
(e.g. \citealt{katz94, sarietal96, tavani96, chiangdermer99}), where the kinetic energy of the flow is 
dissipated via shocks and the prompt emission is the result of synchrotron radiation of relativistic electrons.
The difficulties that this scenario has in dealing with several issues, such as the low-energy photon index \citep{crider97, preece98, preece00} and the 
fine tuning of parameters required to explain the MeV peak of GRB spectra (see also \citealt{beloborodov10} for relevant discussion), motivated works on alternative scenarios. 
For example, variants of non-thermal emission models were discussed in order to overcome the `line-of-death' problem: (i)
effects of adiabatic and/or inverse Compton (IC) cooling in the Klein-Nishina regime on the low energy part of the synchrotron spectrum 
\citep{derishev01, wang09, daigne11}; (ii) synchrotron emission from an electron distribution with a smooth low energy cutoff and
an anistropic distribution in pitch angles \citep{Lloydpetrosian00}; (iii)  the emission observed in the BATSE energy band 20~keV-1~MeV
was interpeted as the result of IC scattering  of slow cooling electrons on the self-absorbed part of the synchrotron spectrum \citep{panaitescumeszaros00};
(iv) synchrotron self-Compton emission under the assumption of continuous electron acceleration \citep{sternpoutanen04}; 
(v) jitter radiation emitted by relativistic electrons moving in non uniform small scale magnetic fields \citep{medvedev00}; (vi) synchrotron electron cooling
in a decaying magnetic field \citep{peerzhang06}; (vii) gamma-ray emission through the Compton-drag process \citep{lazzati00}.
The so-called photospheric models, where the radiation
is released when the outflow becomes transparent, constitute an interesting alternative
to the non-thermal ones. If the energy is dissipated at the very inner parts of the outflow, it thermalizes and
the radiation that escapes from the GRB photosphere has a quasi-thermal
spectrum that peaks at $\sim 0.1-1$~MeV (e.g. \citealt{goodman86, thompson94, beloborodov10}). In the presence
of continuous energy dissipation, however, the resulting spectra may obtain a non-thermal appearance via Comptonization
of the quasi-thermal emission by thermal electrons \citep{meszarosrees00, pe'er06, giannios06, giannios12}.

Hadronic models for the GRB prompt emission constitute a viable alternative. 
 They are built upon the common basis that the (sub)MeV $\gamma$-ray 
emission, which serves as the target field for 
photopion interactions, is not of hadronic origin but it is either the synchrotron radiation of primary electrons 
(e.g. \citealt{dermeratoyan03, asanoinoue07, murase08}) or the emission from the photosphere itself (e.g. \citealt{gao12, asanomeszaros13}). 
The high-energy part
of the gamma-ray spectrum ($>$100~MeV) is typically explained by relativistic proton synchrotron radiation \citep{vietri97,totani98} or by proton-induced
cascades \citep{dermeratoyan06, asanoinoue07,asanoetal09}. The latter scenario has been applied to explain  the
underlying power-law components seen in some bright {\sl Fermi} bursts (e.g. GRB 090902B \citep{abdo09}; GRB 080319B \citep{racusin08}), which
extend from the hard X-rays up to GeV energies and do not agree with simple extrapolations of the MeV spectrum \citep{asanoinoue10}.
In any case, the suggestion that  GRBs are the sources of  ultra-high energy cosmic rays (UHECRs) \citep{waxman95, vietri95, muraseetal08} makes
hadronic models attractive. Moreover, the associated high-energy neutrino emission has been calculated in various studies 
\citep{waxmanbahcall97, murase08, gao12, heetal12, zhangkumar13, asanomeszaros14, baerwald14, reynoso14, petroetal14} and now starts becoming testable 
by ongoing observations \citep{icecube13, aartsen14a,aartsen14b}.

However, one-zone hadronic models are inherently more complex than pure leptonic ones, since they require 
modelling of the coupled emission and energy loss processes (see e.g. \citealt{DMPR12}) between various species in order
to track the evolution of the different components (protons, neutrons, pairs, mesons, neutrinos, photons). 
Several of these feedback processes were proven to give rise to radiative instabilities (e.g. \citealt{sternsvensson91, kirkmast92, mastetal05}) that 
share a common feature: the  abrupt,  i.e. in a few dynamical times, release of energy that is
initially stored in protons and is subsequently transfered to photons. 
The proton synchrotron pair-production instability for example was proposed to give 
rise to gamma-ray emission peaking at $\sim 1$~MeV (e.g. \citealt{kazanas02, mastkazanas06})  offering, at the same time, a physical
connection between the prompt and afterglow phases \citep{mastkazanas09, sultana13}. Moreover,
possible implications of
the automatic $\gamma$-ray quenching instability \citep{stawarz07,petromast11} 
were studied in \cite{petromast12, petromast12b} in the context of hadronic blazar emission.

In the present work we extend this analysis by exploring its emission signatures in gamma-rays, high-energy neutrinos and cosmic-rays
for parameters relevant to GRB sources. We begin with the sole assumption of a 
source that is an efficient ultra high-energy (UHE) proton accelerator 
and sufficiently magnetized in order to confine the accelerated protons.
 Contrary to the majority of GRB hadronic models we do not assume an external source of photons.
We show that for low values of the proton injection luminosity, the only photon field present
in the source is the one emitted by the proton component mainly through synchrotron radiation. 
In this case, the emitted spectrum cannot be assigned to that of a typical
GRB because of its spectral shape and its low luminosity.  We follow the evolution of protons by balancing their losses to the
respective gains of their secondaries as done in \cite{mastkirk95} and show that
if their injection luminosity exceeds a critical value, the system undergoes
a transition that is triggered by the instability of automatic $\gamma$-ray quenching.
The transition is easily identified by an abrupt increase of the photon luminosity 
that causes the source to enter in a high  photon compactness state making the
energy exchange between leptons and photons dominant. This results in photon spectral shapes that
in general resemble GRBs in the sense that (i) they match the required luminosity
and (ii) they can be fit by a Band function. 
It is this self-consistently produced
radiation field that becomes the target for photopion interactions and efficiently 
drains energy from UHE protons, part of which is transfered to high-energy electron and muon neutrinos produced through the charged pion 
decay. This constitutes one of the fundamental differences  between the present study and  others where an {\sl ad hoc} Band spectrum is assumed (see also
\citealt{muraseetal08} for a similar approach to ours). 
Finally, relativistic neutrons produced from the same photopion interactions provide an effective means
for UHECR escape from the source, since they are not magnetically confined and their decay time is long enough as to
allow them to escape freely before converting into protons through $\beta$ decay. 
We show that for modest values of the bulk Lorentz factor ($\Gamma \simeq 100-600$) all three components, namely neutrinos, UHECRs and photons, are energetically similar.

The present work is structured as follows. In \S2 we present the physical conditions of the model and 
derive an analytical expression for the critical proton injection luminosity. 
In \S3.1 we present the numerical code and in \S3.2 we continue with a presentation
of the photon, UHECR and neutrino emission spectra; we also discuss the effects of a lower value for the high-energy cutoff
of the proton distribution. We discuss our results in \S4 and conclude with a summary in \S5.
Throughout this study we use $H_0=70$~km Mpc$^{-1}$ s$^{-1}$, $\Omega_{\rm M}=0.3$, $\Omega_{\Lambda}=0.7$
and $z=1.5$ as an indicative value for the redshift of the gamma-ray source. We also introduce the notation  $Q_{\rm x}=Q/10^x$.

\section{Analytical approach}
\label{analytical}

\subsection{Physical conditions}
The GRB lightcurve in the soft $\gamma$-ray band ($<$10~MeV) is variable and consists of several pulses with durations $\dt$
in the range $0.01-1$~s \citep{norris96, nakarpiran02}. 
Here we adopt the internal shock scenario (see \cite{piran04, zhangmeszaros04}, for reviews) according to which the emitting
region that corresponds to each individual pulse is modelled as a homogeneous shell with Lorentz factor $\Gamma$ 
that forms at a distance  $r \simeq \Gamma^2 c\dt$  from the central engine. 
In the comoving frame the width of the shell is 
\eqb
\rb \simeq r/\Gamma = 3\times10^{11} \Gamma_2 \dt_{-1} \ \textrm{cm}.
\label{rb}
\eqe
As long as
the beaming angle $1/\Gamma$ is smaller than the opening angle of the jet, which
holds during the internal shock phase, we may treat the emission region as
a spherical blob of radius $\rb$. 

Since the typical isotropic energy emitted in $\gamma$-rays 
is $E_{\gamma}^{\rm iso} \simeq 10^{52}-10^{54}$~erg \citep{bloomfrail03,kocevski08}, 
the isotropic $\gamma$-ray luminosity defined as $L_{\gamma}^{\rm iso}\simeq E_{\gamma}^{\rm iso}/ \Delta t $ ranges
between $10^{51}$ and $10^{53}$~erg/s, for a fiducial burst duration $\Delta t=10$~s.
A minimal requirement is that  $L_{\rm tot}\gtrsim L_{\gamma}^{\rm iso}$, where  
where $L_{\rm tot}$ is the total power of the jet and equals to $L_{\rm tot}=L_{\rm k}+L_{\rm B}+\Lpinj$ with
$L_{\rm k}$, $L_{\rm B}$ and $\Lpinj$ being the kinetic, Poynting and  proton\footnote{$\Lpinj$ refers to the luminosity of a power-law proton
distribution.} luminosities, respectively. Although electron acceleration at high energies
is expected to take place too, here, in our attempt to keep the number of free parameters
as low as possible, we assume that the injection luminosity of primary relativistic 
electrons is much lower than that of protons, making their contribution to the energetics 
and the overall spectra negligible. 

We introduce next the parameters $\eb=L_{\rm B}/ L_{\rm k}$ and $\ep=\Lpinj/L_{\rm k}$. We use throughout the text 
$\eb=0.1$ and $\ep=1$ as indicative values, unless stated otherwise.
Using the definition of the Poynting luminosity
\eqb
L_{\rm B} = c B^2 \Gamma^2 r^2
\label{Lpoy}
\eqe
we may write the magnetic field strength measured in the comoving frame as follows:
\eqb
B=\left(\frac{\eb \Lk}{c}\right)^{1/2} \frac{1}{c\Gamma^3 \dt} = 6\times 10^4 \frac{\left(\epsilon_{\rm B, -1} L_{\rm k, 52}\right)^{1/2}}{\dt_{-1} \Gamma_2^3}\ \textrm{G}.
\label{B}
\eqe
As we show next, 
all physical quantities in addition to $\rb$ and $B$ may be expressed through five essential variables: $\Gamma, L_{\rm k},\dt $, $\ep$ and $\eb$.

We proceed with the derivation of 
 the respective expressions for the two basic quantities that describe the proton distribution, namely its injection compactness and
its high-energy cutoff. The former, 
is defined as
\eqb
\lp = \frac{\sth \Lpinj}{4 \pi \rb \Gamma^4 \mpr c^3},
\eqe
and using eq.~(\ref{rb}) it is also written as
\eqb
\lp = \frac{\ep \Lk \sth}{4\pi \mpr c^4 \dt \Gamma^5} = 0.43 \frac{\epsilon_{\rm p, 0} L_{\rm k, 52}}{\dt_{-1}\Gamma_2^5}.
\label{lp}
\eqe
We assume that protons are being injected into the blob after having been accelerated 
into a power-law distribution with index $p_p$ starting from $\gamma_{\min}$ up to 
a Lorentz factor $\gsat$ which is usually determined by the balance between the acceleration
and the energy loss  processes. 
Because protons do not, in principal, suffer as severe radiative losses as electrons do, 
the acceleration of protons to large $\gsat$ in GRBs is possible (e.g. \citealt{asanoetal09, murase12}).
For the lower cutoff of the proton distribution 
and the power-law index we use the indicative values $\gamma_{\min}=1$ and $p_p=2$ respectively.
We note that the exact value of $\gamma_{\min}$ does not alter the main results 
of this work, while steeper proton spectra would
increase the energy demands; for this reason we will not consider such cases here.
Assuming that the synchrotron losses are the dominant energy loss process for high energy protons (see also \citealt{petroetal14})
and that the acceleration process operates close to the Bohm diffusion limit (e.g. \citealt{giannios10}), the typical energy loss and acceleration timescales
are given respectively by $t_{\rm syn} = 6\pi \mpr c \chi^2/\sth B^2 \gamma$ and $t_{\rm acc} = \eta \mpr c^2 \gamma/eBc$, where $\chi=\mpr/\mel$ and $\eta \ge 1$.
Using the above expressions and eq.~(\ref{B}) we find
\eqb
\gsat \simeq 10^9 \Gamma_2^{3/2} \eta_0^{-1/2} \dt_{-1}^{1/2} L_{\rm k, 52}^{-1/4} \epsilon_{\rm B, -1}^{-1/4}.
\label{saturation}
\eqe
Thus, protons can in principle be accelerated to ultra-high energies (UHE), i.e. $E_{\rm p} \le 10^{18}$~eV in the comoving frame. 
If the gyroradius of these highly energetic protons is, however, larger than the typical size of the emission region $\rb$,
they cannot be confined and escape from it. 
Thus, the maximum energy of protons in the emission region is given by $\gmx=\min(\gsat, \gh)$, where
$\gh$ is derived using the Hillas criterion \citep{hillas84}, namely $\gh \mpr c^2 / eB = \rb$. 
Using eqs.~(\ref{rb}) and (\ref{B}) this is also written as 
\eqb
\gh = 6.4 \times 10^9 \left(L_{\rm k, 52} \epsilon_{\rm B,-1}\right)^{1/2}\Gamma_2^{-2}.
\label{hillas}
\eqe
Combining eqs.~(\ref{saturation}) and (\ref{hillas}) we find that $\gmx=\gh$, unless 
\eqb
\Gamma \lesssim 140 \dt_{-1}^{-1/7} \eta_0^{1/7} \epsilon_{\rm B, -1}^{3/14} L_{\rm k, 52}^{3/14}.
\eqe
For the fiducial parameter values used here and for the purposes of the analytical treatment presented in \S 2.2. 
it is sufficient to define the maximum proton Lorentz factor through eq.~(\ref{hillas}).
However, in \S\ref{numerical} where we  study the problem numerically 
for different parameter sets we  use the appropriate expression for $\gmx$.

\subsection{Transition to supercriticality}
The injected protons will emit synchrotron radiation that peaks in the comoving frame  
at $\emx \simeq b \mel c^2 \gamma_{\rm M}^2/\chi$, where
$b=B/\Bcr$, $\Bcr=4.4\times10^{13}$~G, and $\gamma_{\rm M}=\min(\gmx, \gc)$; here, $\gc$ is the typical Lorentz factor of protons that cool due to synchrotron losses 
within the dynamical timescale $\tcr \simeq \rb/c$ and it is given by
\eqb
\gc = 10^8 \frac{\Gamma_2^5 \dt_{-1}}{\epsilon_{\rm B, -1} L_{\rm k, 52}}.
\label{gc}
\eqe
Using eqs.~(\ref{B}) and (\ref{hillas}) we find that 
\eqb
\emx \simeq 12 \frac{\epsilon_{\rm B,-1}^{3/2} L_{\rm k, 52}^{3/2}}{\Gamma^7_{2}\dt_{-1}} \ \textrm{TeV},
\label{quench}
\eqe 
where for simplicity we assumed $\gmx < \gc$ (for the validity of the assumption, see Appendix~A). Thus, for the fiducial parameter values used here 
the peak of the proton synchrotron spectrum falls well within the $\gamma$-ray energy band. 
As long as 
\eqb
\emx & > &  \epsilon_{\rm q}= 0.9 \epsilon_{\rm B, -1}^{-1/6} L_{\rm k, 52}^{-1/6} \Gamma_2 \dt_{-1}^{1/3} \ {\rm GeV}, \ {\rm or} \\ 
\Gamma&  \lesssim  & \Gamma_{\rm q}=300 \ \dt_{-1}^{-1/6} \epsilon_{\rm B, -1}^{5/24} L_{\rm k, 52}^{5/24}
\label{Gq}
\eqe
is satisfied\footnote{For the derivation we used the {\sl feedback} criterion of automatic photon quenching -- see \cite{petromast11}  and eq.~(2) therein.}, 
photons at the high-energy part of the proton synchrotron spectrum can be successible to the instability 
of spontaneous gamma-ray quenching \citep{stawarz07, petromast11}. According to this, the $\gamma$-ray compactness ($\lph$) produced
in a highly magnetized region cannot become arbitrarily high. Whenever it exceeds a critical value ($\lcr$) it is spontaneously absorbed producing
relativistic pairs that cool by emitting a large\footnote{An electron with energy $E_{\rm e} = \mel \gamma c^2$ will emit $N$ photons, where
$N \simeq E_{\rm e} / (\mel c^2 b \gamma^2) = 1/b\gamma = 10^4 / B_{4.6} \gamma_5$.}
number of synchrotron photons, thus providing more targets for further $\gamma \gamma$ absorption.
As the $\gamma$-ray compactess in our framework is related to the proton injection compactess, the existence of an upper limit to $\lph$
is also translated to a limiting value for the proton injection compactess ($\lpcr$).
This has been already pointed out by \cite{petromast12}, although in
a different context.

We refer the reader to Appendix~A for the derivation of $\lpcr$ and here we present the final result
\eqb
 \lpcr  =  \frac{\left(2 \times 10^{-5}\right)\Gamma_2^2}{ \epsilon_{\rm B,-1}^{1/2} L_{\rm k, 52}^{1/2}} \left\{ 
\begin{array} {cl}
22 + \ln \left( \frac{\epsilon_{\rm B, -1}^{1/2} L_{\rm  k, 52}^{1/2}}{\Gamma_2^2}\right), & \gamma_{\rm M}=\gh \\
18 + \ln \left(\frac{\Gamma_2^5 \dt_{-1}} {\epsilon_{\rm B, -1} L_{\rm  k, 52}}\right), & \gamma_{\rm M}=\gc,
\end{array}
\right.
\label{lpcr}
\eqe
where the first branch is relevant for $\Gamma \gtrsim 180 \ \dt_{-1}^{-1/7}\epsilon_{\rm B, -1}^{3/14} L_{\rm k, 52}^{3/14}$ and the second otherwise. 
From this point on we will refer to cases with $\lp < \lpcr$ as `subcritical' and `supercritical' otherwise. We will also drop the logarithmic dependance
and set instead the numerical factor in the bracket of  eq.~(\ref{lpcr}) to 20, i.e. $\lpcr=4\times10^{-4}\Gamma_2^2 \epsilon_{\rm B, -1}^{-1/2} 
L_{\rm k, 52}^{-1/2}$.

\begin{figure}
\centering
\includegraphics[width=8.5cm]{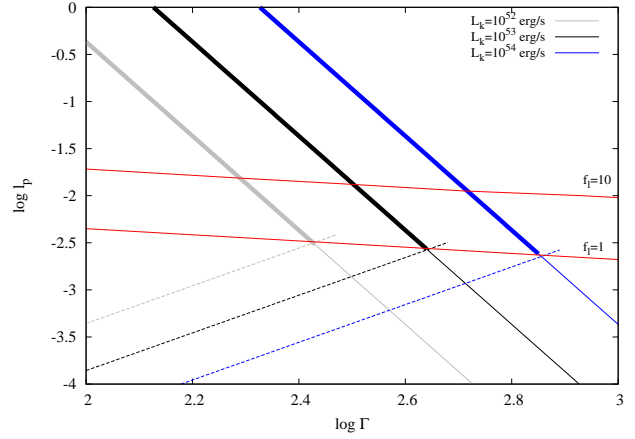}
\caption{Proton injection compactness (solid lines) and proton critical compactness (dashed lines)
as a function of the Lorentz factor for three values of $L_{\rm k}$  marked on the plot. 
Thick lines denote $\lp >\lpcr$, while the loci of points with $f_{\ell}=1$ and 10 are plotted with red lines.
Other parameters used are: $\ep=1$, $\eb=0.1$ and $\dt=0.1$~s.}
\label{param}
\end{figure}

Note that the above expression is valid as long as the gamma-rays that are spontaneously absorbed and initiate
the instability are the result of proton synchrotron radiation. 
If the parameters are such as to push the peak of the proton
synchrotron radiation to GeV energies, e.g. $\eb \ll 0.1$, then the emission 
from pairs produced by Bethe-Heitler and/or photopion interactions of protons with their
own synchrotron radiation dominates in TeV energies. 
It can be shown that even  in such cases the instability can still operate \citep{petromast12}. 
Although the derivation of an expression similar  to eq.~(\ref{lpcr}) 
in this case is out of the scope of the present study, we will present a detailed numerical example in \S\ref{numerical}.


The absolute value of $\lpcr$ is not so important by itself. More important for the evolution of the system
is the ratio $\fl=\lp/\lpcr$ which measures how deep in the supercritical regime the system is driven for given physical conditions.
Using eqs.~(\ref{lp}) and (\ref{lpcr}) this is written
\eqb
\fl = 10^3 \frac{\epsilon_{\rm B, -1}^{1/2} \epsilon_{\rm p, 0} L_{\rm k, 52}^{3/2}}{\Gamma_{2}^7\dt_{-1}},
\label{ratio}
\eqe
where we used the first branch of eq.~(\ref{lpcr}) for simplifying reasons. In any case, both expressions of $\lpcr$ are
similar. Interestingly,  eq.~(\ref{ratio}) shows that 
the condition $\fl>1$ is satisfied for a wide range of parameter values, with lower values of $\Gamma$ being 
prefered. Because of the strong dependance of the ratio $\fl$ on $\Gamma$, slightly different values of the bulk Lorentz factor 
lead to very different photon and neutrino spectra as we show in \S\ref{numerical}.
It is useful, therefore, to define a `critical' value of the Lorentz factor too. Setting $\fl=1$ we find
\eqb
\Gamma_{\rm cr} \simeq  270 \epsilon_{\rm B, -1}^{1/14}  \epsilon_{\rm p, 0}^{1/7} L_{\rm k , 52}^{3/14} \dt_{-1}^{-1/7} ,
\label{gcrit}
\eqe
which has very weak dependance on the parameters. 

 The above are summarized in Fig.~\ref{param} where we plot 
$\lp$ (solid lines) and $\lpcr$  (dashed lines)
for 
three values of $L_{\rm k}$ marked on the plot. Other parameters used are: $\ep=1$, $\eb=0.1$ and $\dt=0.1$~s. 
The dashed lines are plotted up to $\Gamma_{\rm q}$  (see eq.~(\ref{Gq})), as  the derived expression for the
critical compactness is not relevant for larger values of $\Gamma$. 
For each value of $L_{\rm k}$ we plotted with thick lines the part of the $\lp$ curve that lies above
$\lpcr$. Finally, the red lines denote the loci of points with $f_{\ell}=1$ and 10. The parameter space that leads to 
supercriticality becomes wider as $L_{\rm k}$  increases. For high enough values, e.g. $\Lk=10^{54}$~erg/s, supercriticality is ensured
for almost all values of $\Gamma$ relevant to GRBs.
Note also that low values of the bulk Lorentz factor not only favour the transition to supercriticality but also
correspond to $f_{\ell} \gg 10$.

The condition $\fl \gg 1$ implies that the 
system lies deep in the supercritical regime and, as we will show in the next section with detailed numerical examples, 
this has the following implications:
(i) the neutrino production efficiency is high, (ii)
UHE protons cool down effectively through photopair and photopion processes, and (iii) the photon spectrum 
 may be adequately described by a Band function \citep{band93,band09}.

\subsection{Comparison with the WB model}
\begin{figure}
\centering
\includegraphics[width=8.5cm]{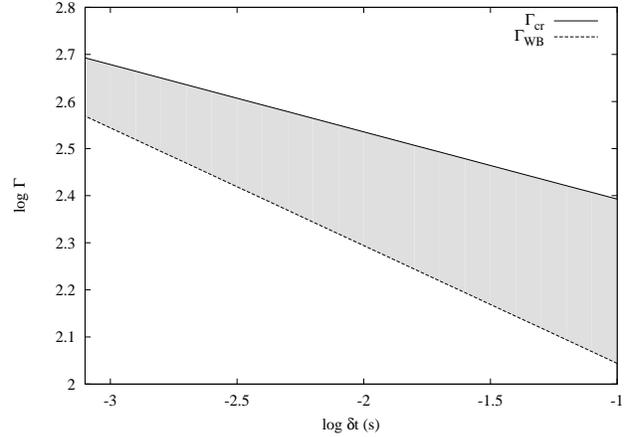}
\caption{$\Gamma$-$\dt$ plane and the two characteristic Lorentz factors 
derived by our analysis (solid line) and the analysis of \citealt{waxmanbahcall97} (dashed line).
The grey colored region denotes additional parameter space with respect to the WB model, where efficient pion production can occur due to 
non-linear feedback loops. For the rest of the parameters used, see text.}
\label{comp}
\end{figure}

Waxman and Bahcall derived an elegant expression for the energy lost by protons through pion production within a dynamical timescale (see eq.~(4) in \citealt{waxmanbahcall97}),
which simply depends on the observed peak energy and $\gamma$-ray luminosity as well as on $\Gamma$ and $\dt$. Since proton cooling due to photopion interactions
becomes more inefficient for larger values of $\Gamma$, their expression can also be seen as an upper limit for $\Gamma$
\eqb
\Gamma_{\rm WB} = 110 \left(\frac{L_{\gamma, 51}}{\epsilon_{\rm obs, 1MeV} \dt_{-1}} \right)^{1/4}.
\label{gWB}
\eqe
In other words, in the WB model, pion and neutrino production is efficient for $\Gamma < \Gamma_{\rm WB}$, which requires either low values
of the Lorentz factor if $\dt\sim0.1$~s or modest values of $\Gamma$, e.g. $\sim 300$, if the 
$\gamma$-ray variability is extremely fast, i.e. $\dt \sim 1$~ms (see e.g. \citealt{waxmanbahcall97, guetta04,abbasi10}).
In our framework, however, efficient pion production is ensured, even  without the requirement
of an {\sl ab initio} target photon field, for parameters leading to the supercritical regime (see e.g. Fig.~\ref{effi} in \S\ref{numerical}).
In the previous paragraph we showed by analytical menas that the transition to supercriticality occurs for $\Gamma \le \Gamma_{\rm cr}$, where 
$\Gamma_{\rm cr}$ is defined in eq.~(\ref{gcrit}).
A comparison between the two Lorentz factors is shown in Fig.~\ref{comp}, where $\Gamma_{\rm WB}$ and $\Gamma_{\rm cr}$ are plotted with
dashed and solid lines, respectively for $L_{\gamma}=10^{51}$~erg/s, $\epsilon_{\rm obs}=1$~MeV, $L_{\rm k}=10^{52}$~erg/s, $\ep=1$ and $\eb=0.1$.

The grey colored area denotes the parameter space where the transition to supercriticality occurs due to the non-linear feedback loops
leading to efficient pion production. As we show in \S\ref{numerical}, the self-consistently produced gamma-ray spectrum
starts resembling a typical GRB one, for $\fl> 10$ or $\Gamma < 0.7\Gamma_{\rm q}$. Thus, our model is equivalent to the WB model in the sense that
efficient pion production on a Band-like gamma-ray spectrum is ensured for roughly similar parameter values, but with a fundamental difference:
here the gamma-ray spectrum is not assumed {\sl a priori} but is produced self-consistently by a series of processes, which we describe in detail in \S3.2.

\section{Numerical investigation}
\label{numerical}
\subsection{Numerical code}
In the previous section we showed by analytical means that the transition to supercriticality is 
ensured for a wide range of parameter values. To verify this we employ the time-dependent numerical code as presented in \cite{DMPR12} -- hereafter DMPR12,
that follows the evolution of protons, neutrons, secondary pairs, photons and neutrinos by solving the coupled differential equations
that describe the various distributions. The coupling of energy losses and injection introduces a self-consistency in this approach that 
allows the study of the system at various conditions, e.g. in the presence of non-linear electromagnetic (EM) cascades and other feedback loops
(see also \citealt{petromast12b} for a relevant discussion). 
 As a word of caution we stress that the aforementioned loops can be fully 
understood only  if the coupled  kinetic equation approach is used. The often used Monte Carlo techniques are intrinsically
linear and fail to capture complex, non-linear effects such as this. 
While Monte Carlo codes are an excellent tool for the description of the system in the subcritical regime, should the choice of 
parameters drive the system into the supercritical regime the results of such codes may be in error.

We assume that protons are being injected in the source at a constant rate  given by
\eqb
Q_{\rm p} = Q_0 \gamma^{-p_{\rm p}} H(\gamma-\gamma_{\min}) H (\gmx-\gamma) H(\tau),
\eqe
where $p_{\rm p}=2$, $\gamma_{\min}=1$, $\gmx=\min(\gh,\gsat)$, and $\tau$ is the time measured in the comoving frame in $\rb/c$ units.
Protons, as well as secondary particles, are allowed to leave the emission region in an average time $t_{\rm esc}=\rb/c$. This 
may account in an approximate way for the expansion of the source, since
the steady-state particle distributions derived by solving a kinetic equation
containing a physical escape term or an adiabatic loss term are similar.
All particles, primary and secondary, lose energy through various processes. 
Although details can be found in DMPR12, for the sake of completeness, we summarize here the physical processes that are included in the code:
\begin{itemize}
\item proton-photon pair production  (photopair) 
\item proton-photon pion production (photopion)
\item neutron-photon pion production
\item proton synchrotron radiation
\item pion, kaon, muon and electron synchrotron radiation
\item synchrotron self-absorption
\item electron inverse Compton scattering
\item photon-photon pair production
\item electron-positron pair annihilation
\item Compton scattering of photons by cooled pairs
\end{itemize}
Photohadronic interactions are modelled using the results of Monte Carlo simulations. In particular, for Bethe-Heitler
pair production the Monte Carlo results by \cite{protheroe96} were used  (see also \citealt{mastetal05}). Photo-pion interactions
were incorporated in the time-dependent code by using the results of the Monte Carlo event generator SOPHIA \citep{muecke00}.
Synchrotron radiation of charged pions and muons was not included in the version of the code presented in DMPR12 and for 
the exact treatment we refer the reader to \cite{dpm14}. 
Pairs that cool down to Lorentz factors $\gamma \sim 1$ contribute to the Thomson depth and they 
are treated as a separate population. Following \cite{lightmanzd87}, we assume that this population thermalizes at a 
temperature $\Theta \ll 1$, where $\Theta=k T_e / \mel c^2$.
For the pair annihilation and photon downscattering processes we 
followed \cite{coppiblandford90} and \cite{lightmanzd87}, respectively (for more details see \citealt{mastkirk95}).  
 
 The photon escape timescale $t_{\gamma, \rm esc}$, in particular, is modelled as
\eqb
t_{\gamma, \rm esc} = \frac{\rb}{c}\left(1+\frac{1}{3} \tau_{\rm KN}(x) f(x)\right),
\label{tgesc}
\eqe
where $x$ is the photon energy in $\mel c^2$ units and $f$ is a function that equals unity for $x\le 0.1$, it
decreases as $(1-x)/0.9$ for $0.1<x <1$ and it becomes zero for $x\ge 1$. Moreover, 
$\tau_{\rm KN} = \tth \sigma_{\rm KN}(x)/\sth$ and $\sigma_{\rm KN}$ is the Klein-Nishina cross section.
 On the one hand, the above expression takes into account in an approximate way the fact that photons may be `trapped'
in the source for longer than one crossing time, i.e. $t_{\gamma, \rm esc} > \rb/c$ for $\tth \gg 1$. On the other hand, 
eq.~(\ref{tgesc}) does not take into account the effect of the expansion of the source during the photon escape.
If the source expands on a dynamical time, the photon escape time is found to be $t_{\gamma,\rm esc} \sim 2 \rb/c$ \citep{giannios06}.
Note, however, that in the examples shown here $\tth \sim 10$, making thus the effect of expansion
rather modest.  
\subsection{Results}
We investigated in total 10 parameter sets that were divided in two groups, namely A and B. All simulations in groups A are obtained for 
$\Lk=10^{52}$~erg/s, $\eb=0.1$, $\ep=1$ and $\dt=0.1$~s, whereas for group B $\dt=0.01$~s. In each group we performed 5 simulations with Lorentz
factors varying between $10^{2.1}$ and $10^{2.5}$ with a logarithmic step of 0.1.
All other parameters that are used as an input for the numerical code,
i.e. $\rb$, $B$, $\gmx$, and $\lp$, are then derived using eqs.~(\ref{rb}), (\ref{B}), (\ref{saturation})-(\ref{hillas}),
 and (\ref{lp}) respectively. These are summarized in Table~\ref{table0}.
In all cases we let the system reach a steady-state, where the photon and neutrino emission as well as the proton and neutron energy distributions
were then calculated. Although, in most cases, a steady-state is achieved in $\sim 2 \tcr$, 
a time-dependent treatment of the GRB emission that is intrinsically variable, is more adequate and it will be the subject of a future work.
\begin{table*}
 \caption{Parameter values used for the calculation of the photon, neutrino and proton energy spectra shown in Figs.~\ref{sed1} and \ref{sed2}. Other parameters used are:
 $L_{\rm k}=10^{52}$~erg/s, $\ep=1$ and $\eb=0.1$.}
 \begin{threeparttable}
 \begin{tabular}{ccccccc}
  \hline
  \# & $\Gamma$ & $B$~(G) & $\rb$~(cm) &  $\min(\gh, \gsat)$ &  $\lp$ & $\lpcr$\tnote{a}\\
   \hline
   \hline
     \multicolumn{7}{c}{Group A: $\dt=0.1$~s}\\
   \hline
1 &  $10^{2.1}$ & $3\times 10^4$ &  $3.8\times10^{11}$& $1.3\times 10^9$ & $1.2\times10^{-1}$ & $6.9\times10^{-4}$\\
2 & $10^{2.2}$ & $1.5\times 10^4$ &  $4.7\times10^{11}$ & $1.7\times 10^9$ & $3.9\times10^{-2}$ & $10^{-3}$ \\
3 & $10^{2.3}$ & $7.7\times 10^3$ &  $6\times10^{11}$ & $1.5\times 10^9$ & $1.2\times10^{-2}$ & $1.7\times 10^{-3}$ \\
4 & $10^{2.4}$ & $3.8\times 10^3$ &  $7.5\times10^{11}$ & $9.2\times 10^8$ & $3.9\times10^{-3}$ & $2.7\times 10^{-3}$ \\
5 & $10^{2.5}$ & $1.9\times 10^3$ &  $9.5\times10^{11}$ & $5.8\times 10^8$ & $1.2\times10^{-3}$ & $4.4\times 10^{-3}$ \\
\hline
\hline
   \multicolumn{7}{c}{Group B: $\dt=0.01$~s}\\
      \hline
6 &  $10^{2.1}$ & $3\times 10^5$ &  $3.8\times10^{10}$& $3.8\times 10^8$ & $1.2$ & $6.9\times10^{-4}$\\
7 & $10^{2.2}$ & $1.5\times 10^5$ &  $4.7\times10^{10}$ & $5.5\times 10^8$ & $3.9\times10^{-1}$ & $10^{-3}$ \\
8 & $10^{2.3}$ & $7.7\times 10^4$ &  $6\times10^{10}$ & $7.7\times 10^8$ & $1.2\times10^{-1}$ & $1.7\times 10^{-3}$ \\
9 & $10^{2.4}$ & $3.8\times 10^4$ &  $7.5\times10^{10}$ & $9.2\times 10^8$ & $3.9\times10^{-2}$ & $2.7\times 10^{-3}$ \\
10 & $10^{2.5}$ & $1.9\times 10^4$ &  $9.5\times10^{10}$ & $5.8\times 10^8$ & $1.2\times10^{-2}$ & $4.4\times 10^{-3}$   \\   
\hline
 \end{tabular}
  \begin{tablenotes}
 \item[a] It is calculated using the first branch of eq.~(\ref{lpcr}) without the logarithmic dependance.
 \end{tablenotes}
 \end{threeparttable}
 \label{table0}
  \end{table*}

\subsubsection{Emission spectra}
\begin{figure}
\centering
\includegraphics[width=8.5cm]{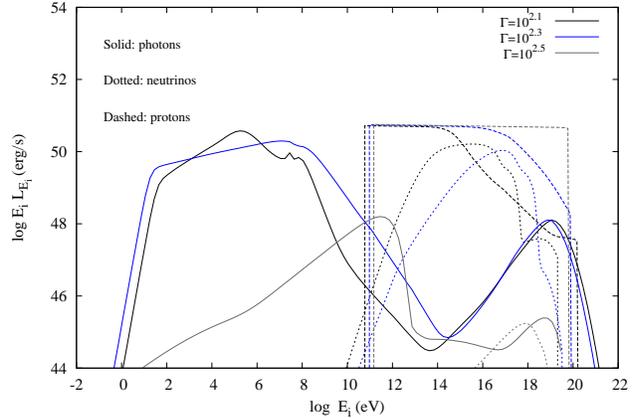}
\caption{Observed $\epsilon L_{\epsilon}$ spectra for a single GRB pulse with duration $\dt=0.1$~s at redshift $z=1.5$. Photon, total neutrino and proton escaping
luminosities are shown with solid, dotted and dashed lines,  respectively, for $\Gamma=10^{2.1}$ (black lines), $10^{2.3}$ (blue lines) and $10^{2.5}$ (grey lines).
The first two cases are supercritical with  $\fl=177$ and 7, whereas the last one is subcritical with $\fl=0.3$. Other parameters
used are: $\Lk=10^{52}$~erg/s, $\eb=0.1$, $\ep=1$.}
\label{sed1}
\end{figure}

Figure~\ref{sed1} shows the observed multiwavelength photon spectra (solid lines) obtained from a single GRB pulse
for $\Lk=10^{52}$~erg/s, $\eb=0.1$, $\ep=1$ and $\dt=0.1$~s (Group A) and for three indicative values of $\Gamma$, i.e. $10^{2.1}$ (black lines), $10^{2.3}$ (blue lines)
and $10^{2.5}$ (black lines). For comparison reasons, the total neutrino\footnote{We refer to the sum of electron/muon neutrino and antineutrino fluxes
as the total neutrino flux.} (dotted lines) and proton escaping (dashed lines) energy spectra are overplotted.
 Since $\Lpinj=\ep \Lk$ is kept fixed, the proton injection compactness 
increases for decreasing $\Gamma$ (see also eq.~(\ref{lp})).
In particular, it increases gradually from $\lp=1.2 \times 10^{-3}$ (lower curve) to $1.2\times 10^{-1}$
(upper curve) with increaments of 0.5 in logarithm. The ratio $\fl$ calculated using eq.~(\ref{ratio}) for each case is given in label of Fig.~\ref{sed1}.

Spectra shown with grey lines are obtained for $\Gamma=10^{2.5}$, which for the particular choice of parameters, leads to
low  $\lp$ and $\fl < 1$. This is a typical example of emission signatures obtained when the system operates in the subcritical regime. 
The photon emission is characterized by the following: the MW spectra are dominated
by the proton synchrotron component, i.e. synchrotron radiation is the dominant energy loss mechanism for UHE protons,
and the radiative efficiency, which is defined as $\eta_{\gamma}=L_{\gamma}/L_{\rm tot}$, is low in agreement with 
typical proton synchrotron emission models (see e.g.~\cite{mueckeprotheroe01}).
In the subcritical regime, the proton spectra at steady state are the same as at injection because of negligible energy losses.
Note that the sharp cutoff of the proton spectra at $\gamma=\gmx$  reflects the injection spectrum $n_{\rm p}\propto \gamma^{-p_{\rm p}}H(\gmx-\gamma)$.
The high energy cutoff would have been smoother if we were to use a more physically motivated injection spectrum, e.g. $n_{\rm p}\propto \gamma^{-p_{\rm p}}e^{-(\gamma/\gmx)^q}$.
The produced neutrinos in this regime are the result of photopion interaction of protons with their own emitted synchrotron radiation (see also DMPR12).
 However, the photopion energy loss rate of protons is very low  and the neutrino emission is supressed. Note that 
 the peak neutrino luminosity is $\simeq 10^{-6}\Lpinj$.

\begin{figure}
 \centering
 \includegraphics[width=6cm]{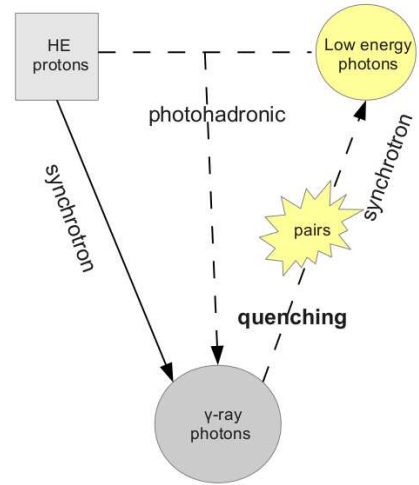}
 \caption{Schematic diagram of the feedback loop that is formed whenever
 spontaneous quenching of proton-produced $\gamma$-rays takes place (dashed lines). 
 }
 \label{loop}
\end{figure}

As the injection compactness increases, at some point it will exceed the critical value (see eq.~(\ref{lpcr})) and 
the system will undergo a {\sl phase} transition, which can be easily identified by a radical change of the spectral shape
and an abrupt increase of the emitted luminosity (see black and blue lines in  Fig.~\ref{sed1}). 
Note that all spectra obtained in the supercritical regime are also characterized by $\fl \gtrsim 1$  in agreement with the analysis of \S2.2.
The underlying reason for this abrupt transition is the instability of spontaneous $\gamma$-ray quenching that
redistributes the energy from the $\gamma$-ray energy band to lower energy parts of the spectrum through an
EM cascade \citep{stawarz07, petromast11}. 
Because of the increased production of low energy photons in the source, it is the photopion/photopair energy loss channels that are favoured over the the proton synchrotron one
while in the supercritical regime.  
 Thus,  the EM cascade initiated by the instability of spontaneous $\gamma$-ray quenching is further supported by
the injection of secondary pairs, which are highly relativistic and
produced by the photopair and photopion interactions of UHE protons with the automatically generated soft photons.
 In other words, once proton-produced $\gamma-$rays reach
a certain compactness they are spontanteously absorbed, giving rise to electron-positron pairs 
and radiation, which causes more proton cooling via photopair and photopion processes; and eventually more $\gamma$-rays, thus sustaining the loop, which is
illustrated in Fig.~\ref{loop}.
\begin{figure}
 \centering
 \includegraphics[width=8.5cm]{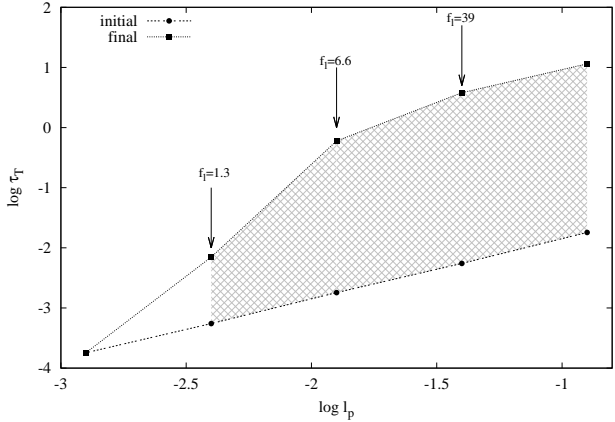}
 \caption{Plot of the Thomson optical depth as a function of $\lp$ in logarithmic scale for the same parameters
 as in Fig.~\ref{sed1}. The values corresponding to the  injection and to the steady state, as derived from the numerical code,
 are shown with circles and squares, respectively. The analytical value of $\fl$ for three values of $\lp$ is also
 shown.}
 \label{tau}
\end{figure}

  All the processes discussed above lead naturally to a large number density of cooled pairs ($n_{\rm e,cool}$) that is related to 
the source's Thomson depth ($\tth$) through $\tth = \sth \rb n_{\rm e, cool}$. 
Since the number of secondaries produced in the EM cascade is related
to the proton injection compactness, we expect that larger values of $\lp$ lead to larger optical depths. 
This trend is exemplified in Fig.~\ref{tau} where we plot the Thomson optical depth versus $\lp$ at the injection (circles) and at the steady state (squares)
for the same parameters as in Fig.~\ref{sed1}. 
The hatched area corresponds to the supercritical regime, while indicative values of $\fl$ (eq.~\ref{ratio}) are marked on the plot for comparison reasons.
At injection we assumed that any electrons present in the source are these related to the injected
protons for conservation of neutrality and, hence, $\tau_{\rm T}^{(0)} = \sth \rb n_{\rm p} \simeq 3 \lp / \ln(\gmx)$. 
At the steady state we find $\tth \gg \tau_{\rm T}^{(0)} $ for $\lp \gg \lpcr$, and this is another manifestation of the increased
secondary pair injection while the system lies in the supercritical regime.

If the value of $\lp$ is such as to result in $\tth > 1$ (blue lines in Fig.~\ref{sed1}), then 
the process that determines the spectral shape is photon Comptonization by cooled electrons.
In particular, our numerical simulations indicate that 
the peak energy of the photon spectrum scales approximately as $\tth^{-2}$, which is in good agreement
with the dependance found by the solving the Kompaneets equation \citep{kompaneets57}. Although similar work can 
be found in the literature (e.g. \cite{arons71, illarionovsunyaev72, lightmanlamb81}), 
we present in Appendix~B an analytic derivation for the case of continuous power-law photon injection
that better describes the physical problem under investigation,  and further supports our numerical results.
Note that if we were to neglect the effects of photon downscattering, then
all spectra obtained in the supercritical regime would have had
the universal shape of a spectrum that  peaks at $\sim 0.1 \Gamma_2$~GeV in the observer's frame (see e.g. blue line in Fig.~\ref{sed1}),
which reminds of the power-law underlying component seen in several bright GRBs (e.g. 080319B, 090902B, 090926A)
detected with {\sl Fermi}  \citep{racusin08, abdo09, ackermann11}. Cascade emission produced
through proton interactions with the MeV photons of the GRB were suggested as an alternative explanation for this emission (e.g. \citealt{asanoinoue10}).
Interestingly, we derive similar photon spectral shapes which are also the result of a cascade but with the key difference
that a photon field is not required {\sl ab initio}; the targets are provided by the quenching loop.
In the succession of spectra shown in Fig.~\ref{sed1} only those obtained for $\Gamma=10^{2.1}$ 
correspond to large enough $\lp$ and $\tth$ to start resembling to a Band-like photon spectrum \citep{band93}. 
The exact spectral shapes, however, should be considered with caution, since 
a better description for the cooling of pairs below $\beta \gamma\sim 1$ is required and which we plan
to address in the future.

As the system is driven deeper to the supercritical regime the  photopion energy losses become gradually more significant and 
the energy drained from UHE protons is transfered to photons and secondary particles, such as pairs and neutrinos. 
The abrupt increase of the neutrino fluence is an additional sign of the transition to the supercritical regime -- see grey and black lines in Fig.~\ref{sed1}.
As the proton injection compactness progressively increases (bottom to top) the neutrino spectrum becomes flatter in $\epsilon F_{\epsilon}$ units,
and it extends to lower energies tracing the evolution of the peak photon luminosity. The neutrino spectra 
for $\lp \gg \lpcr$ share many common features with those obtained in studies where the
photon target field is modelled by an {\sl ab initio} Band spectrum (e.g. \citealt{murase08, baerwald11,baerwald13, petro14}).
\begin{figure}
\centering
\includegraphics[width=8.5cm]{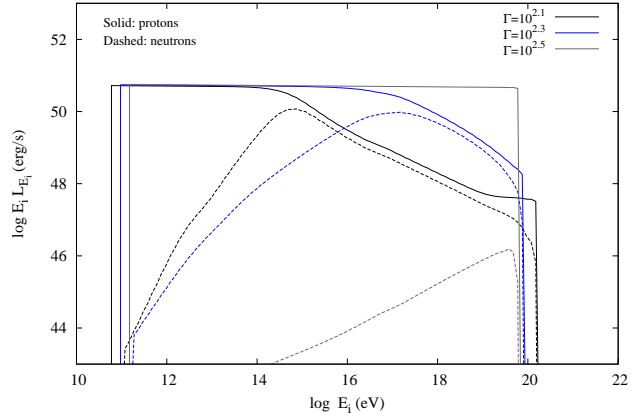}
\caption{Observed $\epsilon L_{\epsilon}$ spectra for protons (solid lines) and neutrons
(dashed lines) for the same parameters as in Fig.~\ref{sed1}.}
\label{pn}
\end{figure}
\begin{figure}
\def\tabularxcolumn#1{m{#1}}
\begin{tabularx}{\linewidth}{@{}cXX@{}}
\begin{tabular}{c}
\subfloat[]{\includegraphics[width=8.5cm]{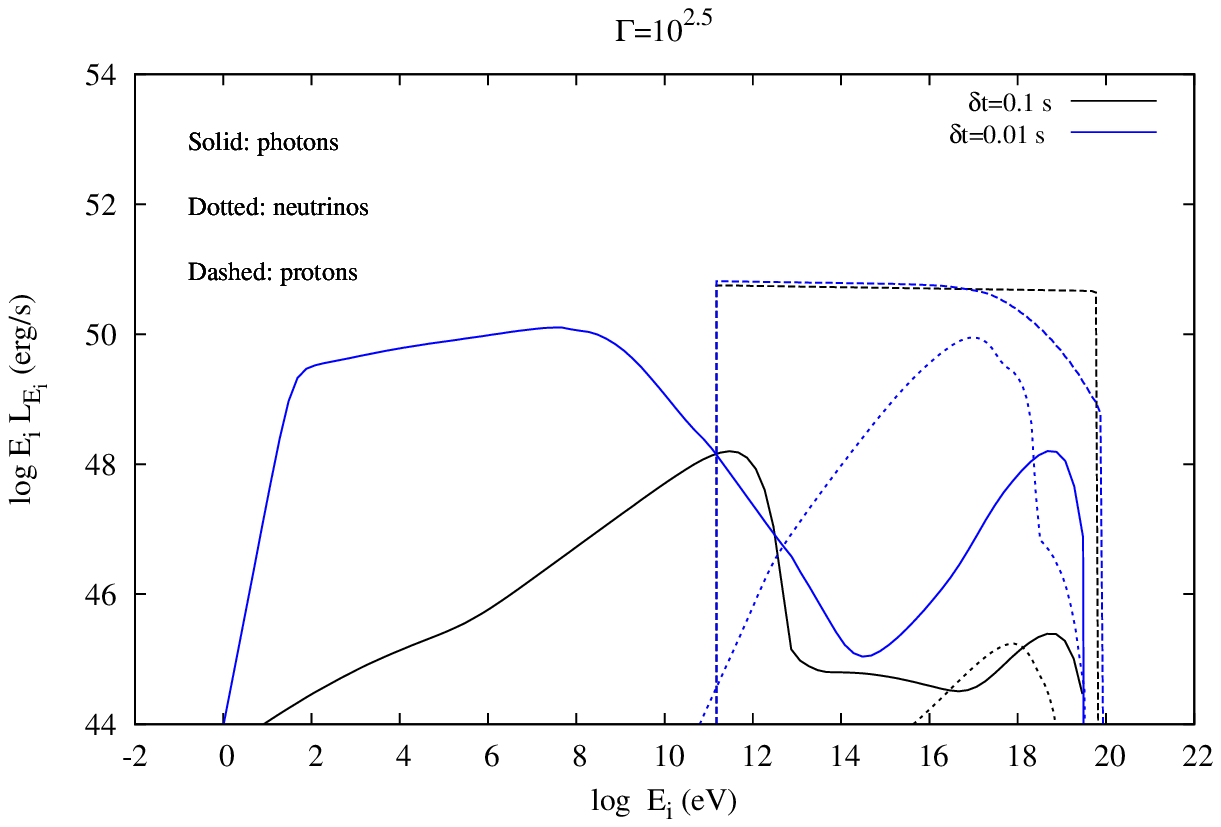}}\\
\subfloat[]{\includegraphics[width=8.5cm]{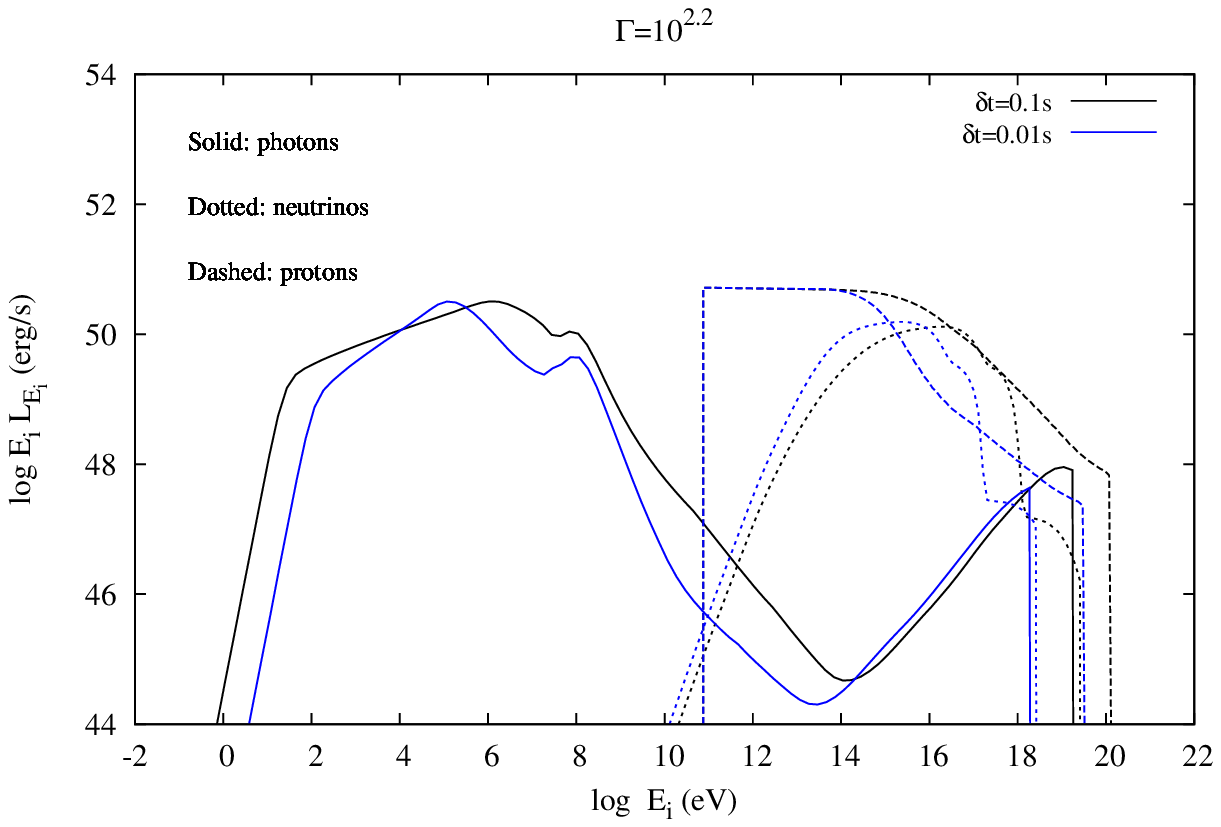}} \\
\end{tabular}
\end{tabularx}
\caption{Comparison of observed photon, neutrino and proton energy spectra for $\dt=0.1$~s and $\dt=0.01$~s shown with black and blue lines, respectively.
Spectra in panels (a) and (b)  are obtained for $\Gamma=10^{2.5}$ and $\Gamma=10^{2.2}$, respectively. All other parameters are same as in Fig.~\ref{sed1}.}
\label{sed2}
\end{figure}
\begin{figure}
\centering
\includegraphics[width=8.5cm]{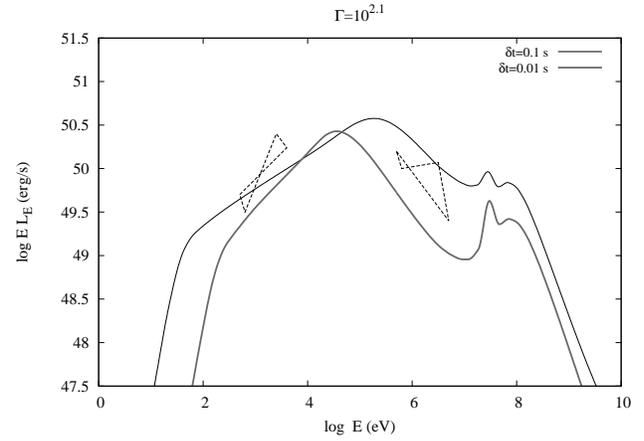}
\caption{Gamma-ray spectra obtained for $\dt=0.1$~s (black lines) and $\dt=0.01$~s (grey lines). Other parameters used are:
$\Lk=10^{52}$~erg/s, $\eb=0.1$, $\ep=1$ and $\Gamma=10^{2.1}$. 
The bowties show the range of observed values for the GRB photon indices, namely $-1.4 \le \alpha \le 0.5$ and $-2.8 \le \beta\le  -1.9$ (\citealt{preece00}),
and are plotted for guiding the eye.
}
\label{band}
\end{figure}
Summarizing, conditions leading to high photon and neutrino luminosities cause unavoidable
cooling of UHE protons (see also \citealt{asano05}) that obtain a steeper power-law spectrum than that of injection, i.e.
$n_{\rm p} \propto \gamma^{-s}$, where $s\approx p_{\rm p}+1/2$.
Figure \ref{pn} compares the observed differential luminosity of escaping protons and neutrons for the same parameters as in Fig.~\ref{sed1}. 
 In all three cases, we find that protons with energies $\lesssim 10^{14}$~eV are unaffected by cooling due to photohadronic processes
and actually serve as a large energy reservoir. However, their contribution to the spectral and temporal properties of the gamma-ray
spectra is negligible both in the subcritical and supercritical regimes, and this
can be understood as follows: protons with $E_{\rm p} < 10^{14}$~eV
emit mainly through synchrotron at low photon energies, e.g. $E_{\gamma}< 2.5 {\rm eV} (E_{\rm, p}/10^{14} {\rm eV}) ^2 
(B/10^4 {\rm G}) (100/\Gamma)$ and with a much lower luminosity than the gamma-ray one\footnote{The gamma-ray luminosity in the subcritical regime
is given by the peak luminosity of the proton synchrotron component, while in the supercritical regime is given by the peak 
of the cascade emission component (see Fig.~\ref{sed1}).}.  In this context it is the high energy 
part of the proton distribution that is `active'.
The luminosity carried by neutrons, which are produced via the photopion channel $p\gamma \rightarrow n \pi^{+}$,
increases as the cooling of UHE protons becomes progressively more significant, i.e. as the conditions
lead the system deeper into the supercritical regime. At the same time, the peak of the neutron energy
spectrum moves towards lower energies, similarly to the cooling break energy of the proton energy spectrum, 
and the neutron distribution can be described by the same  power-law  as cooled protons, i.e., with index $p_{\rm n}\simeq s \approx p_{\rm p}+1/2$.
In the optically thin limit for photopion interactions the produced neutron distribution would follow the proton 
injection spectrum. Here, the steepening of the neutron spectrum is the result of efficient neutron cooling through 
$n\gamma \rightarrow p \pi^{-}$ before escape from the emission region.  Contrary to protons,  neutrons are not confined by magnetic fields and 
their decay time is large enough to allow them to escape freely before converting through $\beta$-decay into
protons. These will propagate as UHECRs into the intergalactic medium having an injection spectrum similar to that of their parent
population \citep{kirkmast89, begelman90, giovanoni90, mannheim01, atoyandermer03}. Other possible escape mechanisms of UHECRs are
discussed in \cite{baerwald13, asanomeszaros14}. 

Having explained the basic features of the photon, neutrino and proton spectra as the system is driven progressively from 
the subcritical to the supercritical regime for a particular parameter set (Group A), we proceed 
to investigate the role of other parameters, such as $\dt$ -- see Fig.~\ref{sed2}.
If all other parameters are kept fixed, faster variability is translated to higher proton injection compactness ($\lp \propto \dt^{-1}$),
higher magnetic field strength ($B\propto \dt^{-1}$), smaller emission region ($\rb \propto \dt$) but approximately constant
$\lpcr$, as the only dependance on $\dt$ comes through the logarithmic term (see eq.~(\ref{lpcr})). Thus, for the same $\Gamma$ but 
smaller $\dt$ the system is driven
more deep to the supercritical regime (panel (a) in Fig.~\ref{sed2}). 
The effect of the stronger magnetic field for $\dt=0.01$~s is also 
imprinted on the cutoff energy of the neutrino spectrum, which moves
from $\sim 100$~PeV to 10~PeV (panel (b)). 
The gamma-ray emission produced for $\Gamma=10^{2.1}$ in both cases is shown in Fig.~\ref{band}. The 
bowties show the range of observed values for the GRB photon indices, namely $-1.4 \le \alpha \le 0.5$ and $-2.8 \le \beta\le  -1.9$ \citep{preece00},
and are plotted for guiding the eye. Note the change
of the gamma-ray spectrum below the peak, which becomes harder as $\lp \gg \lpcr$.

\begin{figure*}
\def\tabularxcolumn#1{m{#1}}
\begin{tabularx}{\linewidth}{@{}cXX@{}}
\begin{tabular}{c c }
\subfloat[]{\includegraphics[width=8.5cm]{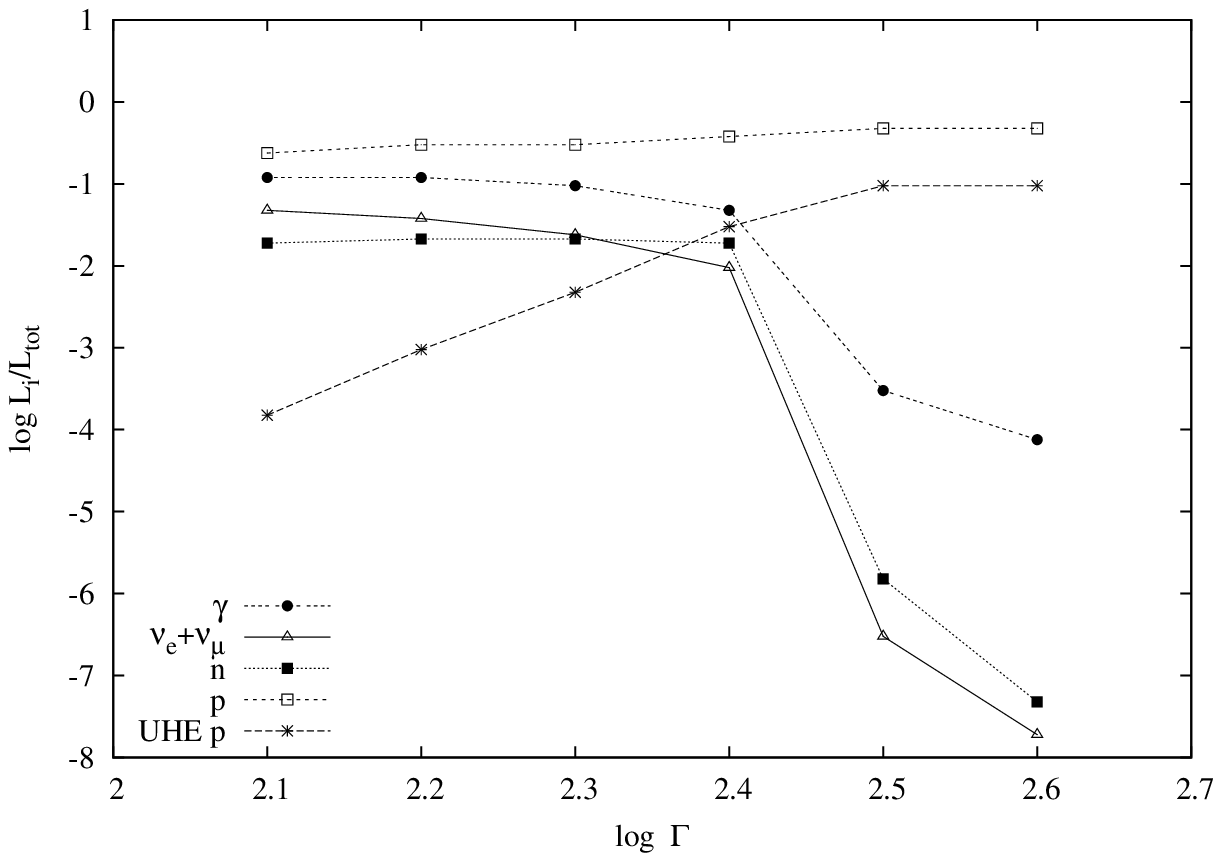}} & 
\subfloat[]{\includegraphics[width=8.5cm]{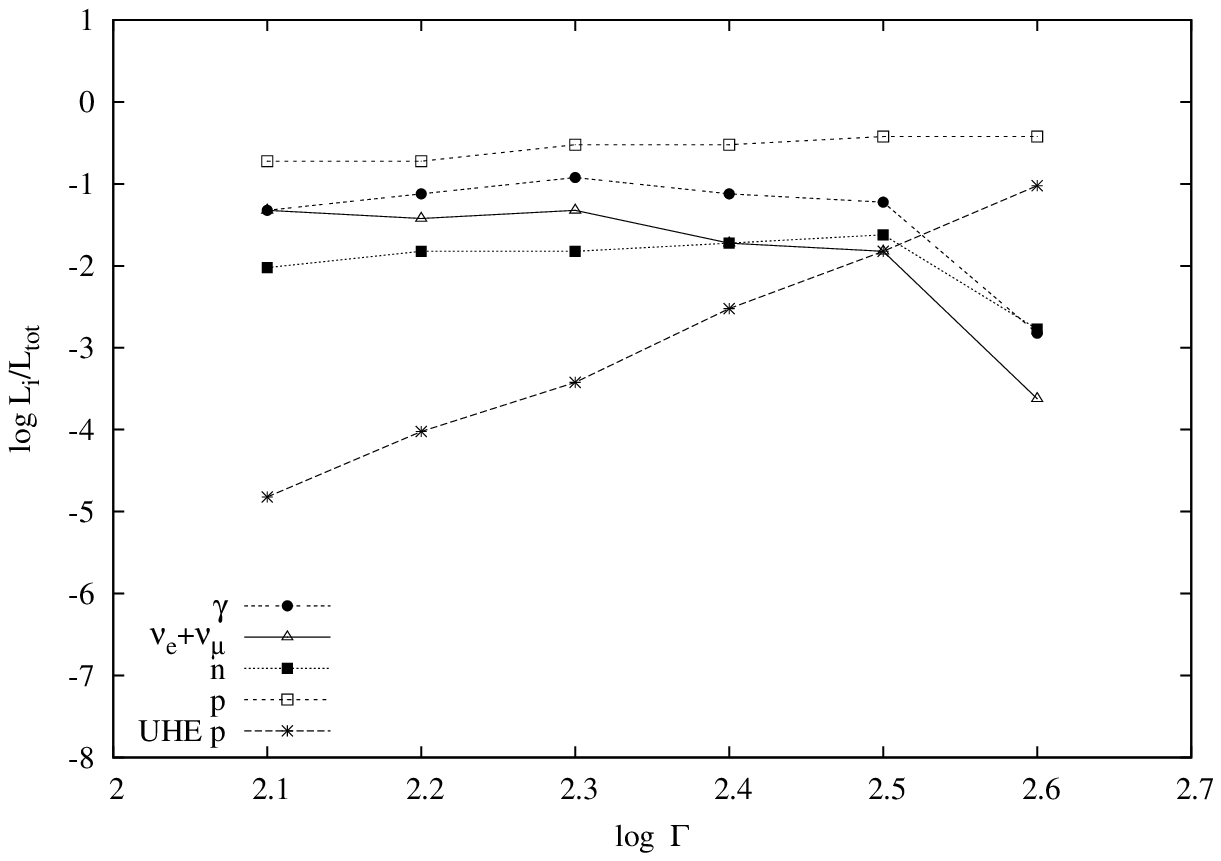}} \\
\end{tabular}
\end{tabularx}
\caption{Log-log plot of the ratio $L_{\rm i}/L_{\rm tot}$ as a function of the Lorentz factor for parameter sets from Group A and B shown
in panels (a) and (b), respectively. The subscript $i$ accounts for photons, electron and muon neutrinos, neutrons, 
protons, and UHE ($>10^{18}$~eV) protons. All parameters are the same as in Figs.~\ref{sed1} and \ref{sed2}.}
\label{effi}
\end{figure*}
\subsubsection{Efficiency}
A robust sign of the transition to supercriticality 
is the abrupt increase of the photon, neutron and neutrino luminosities. 
The efficient conversion of energy originally stored
in relativistic protons into radiation can be the result of other underlying feedback loops, such as the `PPS-loop' \citep{kirkmast92}, which
was also studied in the framework of GRB prompt emission \citep{mastkazanas06,mastkazanas09}.
Here we define the efficiency of the $i$-th component as $\eta_{\rm i}=L_{\rm i}/L_{\rm tot}$, where  $L_{\rm tot}= \Lk(1+\ep+\eb)$, 
$i=\gamma$ (photons), $\nu$ (neutrinos), $n$ (neutrons),
$p$ (protons) and $P$ (protons with energies $>10^{18}$~eV).
The efficiency as a function of the Lorentz factor is shown in  
Fig.~\ref{effi} with panels (a) and (b) corresponding to Groups A and B, respectively. 
Few things are worth commenting:
\begin{itemize}
 \item The luminosity of the proton component is the dominant one. The photon luminosity becomes comparable to the luminosity 
 carried by the proton component only for parameters
 that drive the system deep into the supercritical regime (see e.g. Fig.~\ref{param}), where we 
 typically find that $\eta_{\gamma} \approx 0.1-0.2$ or $L_{\gamma} \approx (0.2-0.4) \Lpinj$.
 \item The abrupt increase of  $\eta_{\gamma}$, $\eta_{\nu}$ and $\eta_{\rm n}$ is seen 
       for $\Gamma \simeq 10^{2.4} \simeq \Gamma_{\rm cr}$ (see also eq.~(\ref{gcrit})) and it marks the transition to supercriticality. 
 \item In panel (b) the abrupt increase of the neutrino, neutron and photon luminosities is not evident, as our simulations do not extend above  $\Gamma_{\rm cr}$, which
        in this case is $\simeq 10^{2.6}$.        
 \item The steep decrease of $\eta_{\rm P}$ ($\sim 3$ orders of magnitude for a 0.3 order of magnitude 
       change in $\Gamma$) indicates the significant cooling of UHE protons.
 \item For $\Gamma < \Gamma_{\rm cr}$ we find $\eta_{\gamma} \gtrsim \eta_{\nu}$. For $\Gamma > \Gamma_{\rm cr}$ on the other hand,
       $\eta_{\nu}$ decreases faster than $\eta_{\gamma}$, since photopion interactions are the sole source of neutrinos contrary to photons,
       which are produced mainly via synchrotron radiation in the subcritical regime.
 \item The efficiencies in neutrons and neutrinos are, generally, of the same order of magnitude. They show, however,
 different trends: $\eta_{\rm n}$ remains approximately constant (panel (a)) or decreases (panel (b))  for smaller values of $\Gamma$, whereas
 $\eta_{\nu}$ increases. This implies that neutrons interact with photons before escaping from the source and contribute
 to the neutrino production through the process $n\gamma \rightarrow p \pi^{-}$.
 \end{itemize}

\subsubsection{Maximum proton energy}
We continue our study on the emission 
features while the system is in the supercritical regime by considering a fiducial case where protons are not accelerated up to 
UHE, i.e. the maximum proton energy in the comoving frame is $\ll 10^{18}$~eV. One could imagine a scenario where
the magnetization of the burst is very low (e.g. $\eb \ll 10^{-3}$)  or the size of the emission region
is small enough to confine higher energy protons.
\begin{figure*}
\def\tabularxcolumn#1{m{#1}}
\begin{tabularx}{\linewidth}{@{}cXX@{}}
\begin{tabular}{c c}
\subfloat[]{\includegraphics[width=8.5cm]{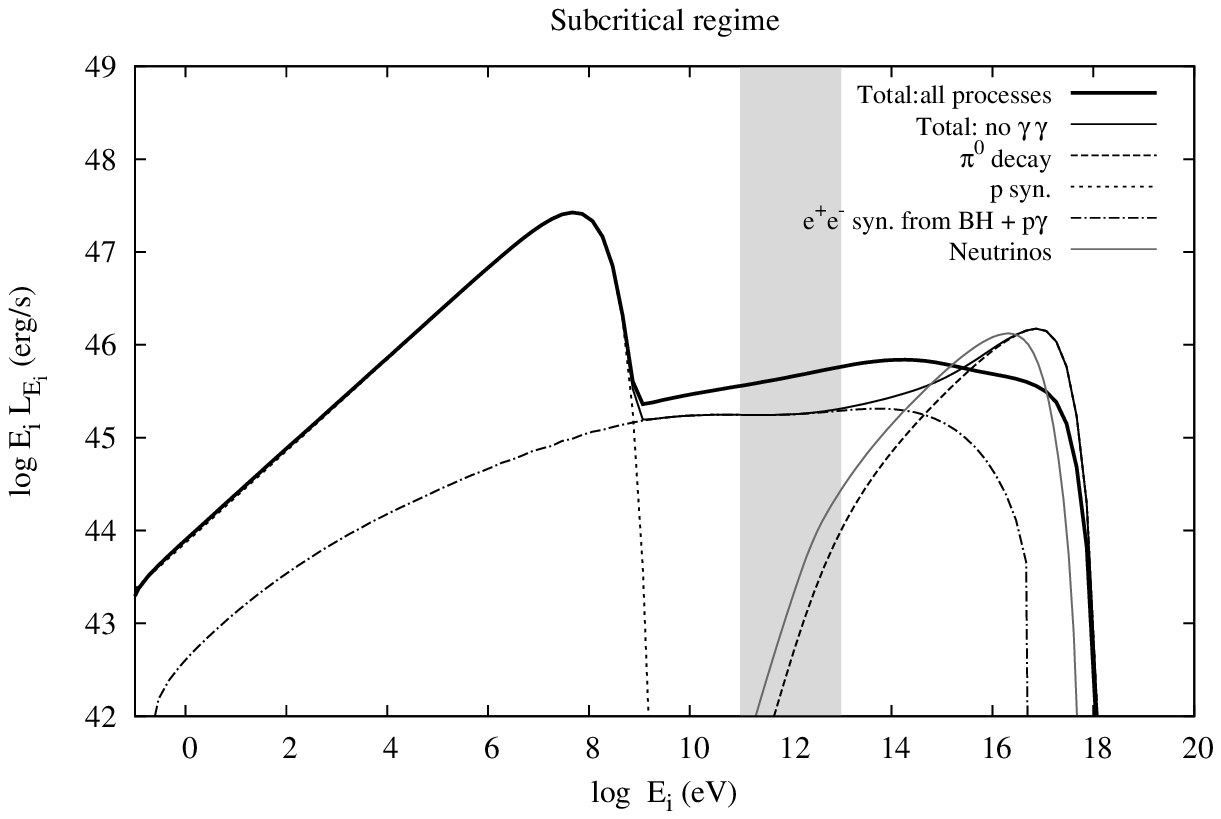}} & 
\subfloat[]{\includegraphics[width=8.5cm]{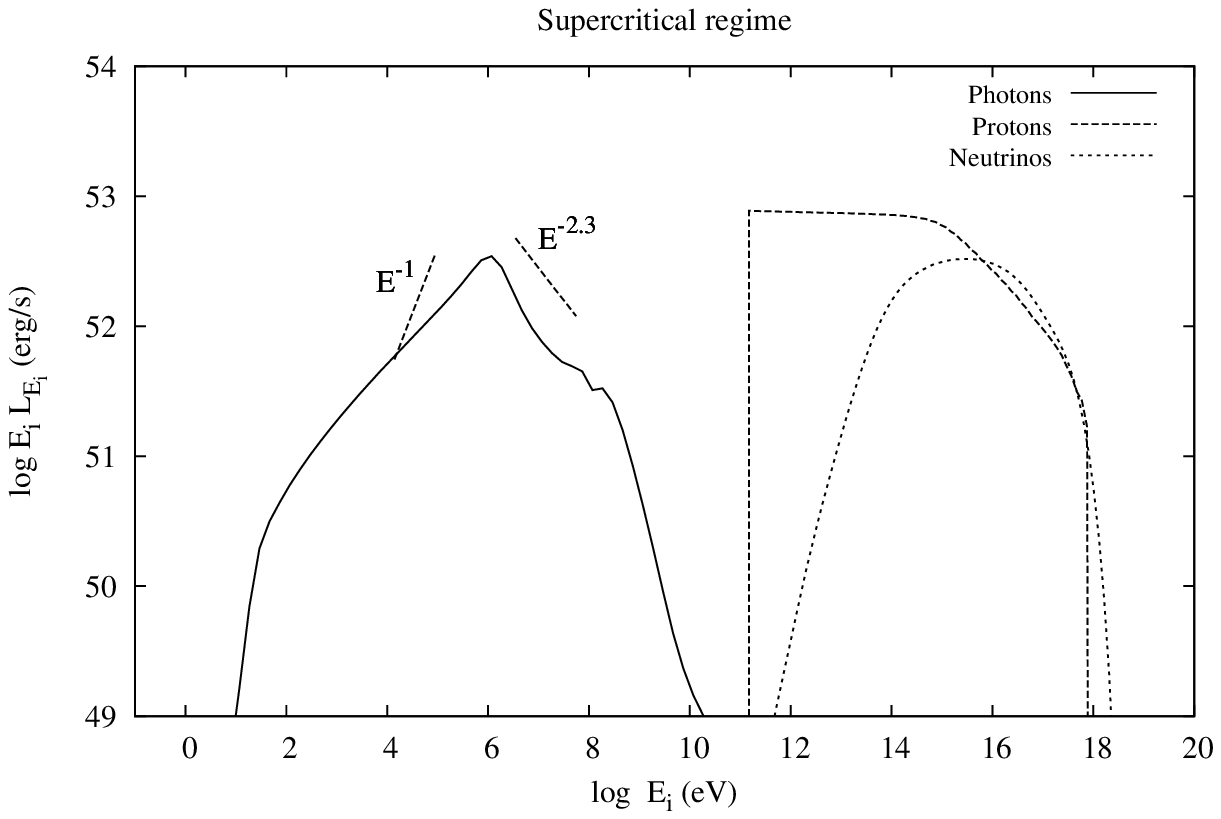}} \\
\end{tabular}
\end{tabularx}
\caption{Examples of multiwavelgth spectra obtained for a lower value of the proton injection energy ($\gmx=7\times 10^6$)
in the subcritical (panel a) and supercritical (panel b) regime. 
Panel (a): The total photon spectrum when all processes are taken into account is plotted with thick black line, whereas
when $\gamma \gamma$ absorption is omitted the result is shown with thin black line. The contribution to the total flux of various
components is also shown: proton synchrotron (dotted line), gamma-rays from $\pi^0$ decay (dashed line) and synchrotron from 
pairs from BH and photopion processes (dashed-dotted line). The neutrino spectra (grey line) are also overplotted. 
The grey colored region marks the 0.1-10~TeV energy band. 
Panel (b): The total photon (solid line), neutrino (dotted line) and proton (dashed line) energy spectra  are shown . The
typical low- and high-energy photon indices are also shown.
Other parameters used are: $L_{\rm k}=10^{53}$~erg/s, 
$\Gamma=10^{2.5}$, $\dt=0.1$~s, $\eb=0.01$, $B=1.9\times10^3$~G and $\rb=9.5\times10^{11}$~cm; the proton injection luminosity
used   in panels (a) and (b) is $10^{53}$~erg/s and $10^{54}$~erg/s, respectively. }
\label{lowgpmax}
\end{figure*}
To exemplify the above we adopt the following parameters: $\gmx=7\times 10^6$, $L_{\rm k}=10^{53}$~erg/s, $\Gamma=10^{2.5}$, $\dt=0.1$~s, $\eb=0.01$ and 
two values of the proton injection luminosity, $\Lpinj=10^{53}$~erg/s and $10^{54}$~erg/s, which correspond to 
the subcritical and supercritical regime, respectively.  Other parameters used are: $B=1.9\times10^3$~G and $\rb=9.5\times10^{11}$~cm. 

The results for the subcritical and supercritical cases are  summarized in panels (a) and (b) of Fig.~\ref{lowgpmax}, respectively.
In the subcritical regime the various components of the overall photon spectrum can be identified, in contrast to 
the supercritical regime where the formation of the EM cascade blurs the emission signatures from individual processes.
In panel (a) we plot the various contributions in order
to demonstrate that in the TeV energy band (grey colored region) the gamma-ray emission is no longer dominated  by the proton synchrotron component (for comparison see
grey line in Fig.~\ref{sed1}). Instead, it is the synchrotron radiation from secondary pairs produced through the Bethe-Heitler and photopion processes 
that is being emitted as Tev gamma-rays and that it is going to initiate the instability of automatic photon quenching (for relevant
discussion see also \citealt{petromast12b}).

Note that the emission signatures of hadronic plasmas in the subcritical regime may differ significantly for
various parameter sets, as we exemplified in Figs.~\ref{sed1} and \ref{lowgpmax}. However, 
the photon and neutrino emission produced when the system is driven to the supercritical regime,
and in particular for parameters which ensure $f_{\ell}\gg 1$, is rather `universal'.
Besides the overall energetics, the photon SEDs shown in Figs.~\ref{sed1} (black line), \ref{band} and \ref{lowgpmax} have similar features.

We also found that for lower values of $\gmx$, a higher value $\lp$ is generally required  to enter the supercritical regime.
Given that in this regime approximetaly $(0.2-0.4)\Lpinj$ goes to gamma-rays and neutrinos (see also Fig.~\ref{effi}), the above conditions lead inevitably
to bright  photon and neutrino bursts. For the case shown in Fig.~\ref{lowgpmax} for example, the gamma-ray fluence emitted in the supercritical regime 
would be $\sim 3\times10^{-5}$ erg/cm$^2$, if we were to assume a duration of $10$~s.
This would place such an event at the high-fluence tail of the Fermi distribution \citep{fermiGBM14}.


\section{Discussion}
Hadronic models have served for a long time as
alternatives to the more popular leptonic ones for AGN and GRB high-energy emission.
One of their  unique features is the prediction of copious neutrino emission which
can be produced alongside the photon spectrum   and is particularly attractive because
it allows for hadronic models to be further tested.
These suffer, however, from low efficiencies, 
as hadrons have in general long cooling timescales.

In earlier works \citep{mastetal05, dpm12,petromast12,petromast12b} that were
targeting AGN high-energy emission, we showed that
we can divide the parameter space of hadronic plasmas
into two regimes. In the first one, which we shall 
call `subcritical', protons carry the 
majority of the energy while a small amount is radiated away mainly by
proton synchrotron radiation; photon induced processes like photopair
and photopion carry even less luminosity -- this is
especially true if one assumes that
there are no ambient photons illuminating the source. In this regime the system
is  inefficient, i.e. protons lose a very small part of their energy
to secondaries. The other regime, the `supercritical' one, 
is separated by a sharp boundary in phase space from the subcritical one 
and is characterised by exactly the opposite
trend. Here, processes like photopair and photopion production dominate the
losses and essentially drain the protons of their stored energy
giving it to secondaries, thus increasing the efficiency to high values.
Because of the radical change not only in the efficiency but also in the emission signatures of such
plasmas, the transition from the sub-to the super-critical regime can be characterised as a `phase' transition, with
the underlying reason being the existence of non-linear feeback loops. 

In the present study we re-examined the radiative signatures of 
hadronic plasmas in the context of GRBs. 
The key question we wanted to answer was whether the energetics required for the supercriticality 
are compatible with typical GRB parameters. 
For this, we  adopted the  usual GRB hadronic picture  used in the literature, i.e. we assumed the injection of  
high energy protons having a power law distribution
in a 
source of a given size that contains a certain magnetic field and  adopted  parameter values relevant
to GRB sources. However, departing from the usual GRB assumptions, neither external photons nor an
extra population of accelerated relativistic electrons were considered.
 This choice minimized the number of free parameters to five: $\eb$, $\epsilon_{\rm p}$, $L_{\rm k}$, $\Gamma$ and $\dt$;
all other quantities, such as the proton injection luminosity ($\Lpinj$)
and the maximum energy ($\gmx$) of their distribution, can be expressed in terms of these parameters
(see eqs.~(\ref{rb}), (\ref{B}), (\ref{lp}), (\ref{saturation}), (\ref{hillas})).
 For the derivation of $\gmx$ we assumed that proton acceleration is fast, i.e. $t_{\rm acc} \simeq t_{\rm g}$, where
$t_{\rm g}$ is the proton gyration timescale. For example, stochastic Fermi acceleration due to magnetic turbulence 
(e.g. \citealt{waxman95b, dermerhumi01}) as
well as relativistic magnetic reconnection (e.g. \citealt{giannios10}) have been suggested as viable processes
for fast, UHE proton acceleration.
Whether or not the accelerated protons can form a power-law distribution 
remains to be shown, although there are some indications for the formation of a high-energy proton power-law tail
in simulations of relativistic reconnection in electron-ion plasmas (private communication with Dr. L.~Sironi). Note
that the same kind of numerical studies performed for pair plasmas clearly show the formation of a power-law electron
distribution \citep{sironispitkovsky14}.

As a tool for our study we employed a recently developed numerical code \citep{DMPR12}
for solving the system of spatially averaged kinetic equations, that describes the coupling between
protons and their stable by-products, namely photons, electrons (and positrons), neutrons and neutrinos.  This, 
 in contrast to most of the work performed on hadronic plasmas thus far, has allowed us to make a self-consistent study
of the evolution of the system by keeping track of the energy lost and gained by the
various species. We have confirmed that this coupling between the species is 
the key to understanding its behaviour as only with this scheme can one follow
the development and growth of non-linear loops that eventually lead the system 
to supercriticality.

Altough there are various loops that may cause the phase transition from sub- to super-criticality (e.g. \citealt{kirkmast92, kazanas02, mastetal05}),
our analysis presented in \S\ref{analytical} and in Appendix~A  shows that $\gamma-$ray quenching \citep{stawarz07, petromast11} 
is the leading one for the parameters used. This means that once proton-produced $\gamma-$rays reach
a certain compactness they are spontanteously absorbed, giving rise to electron-positron pairs 
and radiation, which causes 
more proton cooling via photopair and photopion processes; and eventually more $\gamma$-rays, thus sustaining the loop.
The cycle will continue until protons are  drained
of their energy and this, depending on the initial conditions, can lead the system either
to steady state or to a limit cycle behaviour -- in the latter case energy is gradually built into
protons and is abruptly released in a few crossing times.
Note that all the examples presented in  \S\ref{numerical} were obtained for parameters that led quickly 
(in 1-2 dynamical times) to a steady state.

As we have shown in \S2, it is the existence of a critical $\gamma$-ray compactness that makes 
the proton compactness $\lp$ the most important parameter in our study.
On the one hand, if we use 
a similar definition to the radiation compactness and 
relate  the observed  proton luminosity, which is assumed to be a fraction $\ep$ of the jet's kinetic luminosity $L_{\rm k}$, 
to the one measured in the comoving frame of the flow through a Lorentz transformation, we find $\lp \propto \ep L_{\rm k} \dt^{-1} \Gamma^{-5}$.
On the other hand, we showed that the critical $\gamma$-ray compactness is translated to a critical value for the proton one, which roughly scales as  
$\lpcr \propto \Gamma^2 \eb^{-1/2} L_{\rm k}^{-1/2}$.  The combination of the $\Gamma^{-5}$ and $\Gamma^2$ dependances favours supercriticality 
for most parameter values, except for flows with high bulk Lorentz factors. These results are summarized in Fig.~\ref{param},
which answers in the most satisfactory manner
our earlier posed key question.

Apart from the above, there are 
some  far reaching consequences of this model when applied to GRBs.
First,  the efficiency of photons and neutrinos becomes quite high, reaching 0.1-0.4
of the total available proton luminosity. Moreover, the high number 
of electron-positron pairs created as secondaries cool to low energies  producing a high Thomson
optical depth (ranging from a few up to $\mpr/\mel$) which
can downscatter high energy photons producing a  bump at ~$\mel c^2 \tau_T^{-2}$
in the rest frame of the flow (see Appendix~B).  It is noteworthy that a similar idea for explaining the 
peak of the GRB emission was proposed by \cite{brainerd94}, although in a different context.
Moreover, the boundary in the phase space between the 
two regimes  is very sharp, in the sense that a small perturbation in one of the proton injection parameters,
while the system is still in the subcritical regime, can push it over to the supercritical one.  
The transition is in most cases very abrupt and it manifests itself with a photon
flare which lasts a few crossing times \citep{DMPR12, petromast12b}.

Clearly there are open questions that one needs to address before the present model can 
 successfully explain the GRB phenomenology, such as gamma-ray spectra and time-variability, which 
we plan to investigate in a forthcoming publication. For the former, we plan
to include an additional equation for low-energy electrons ($\beta \gamma < 1$) and follow the cooling/heating due to Compton
process in more detail. Regarding the latter, one of the central issues that has to be addressed
is whether or not our hadronic model can produce the fast variability observed in $\gamma$-rays.
Preliminary results of photon lightcurves derived in the supercritical regime assuming a variable proton injection rate 
show that rapid variability (of the order of the source crossing time) can be obtained. 
Note that even for a constant proton injection rate
there are regimes in the parameter space that are relevant to GRBs
and lead to a limit cycle behaviour \citep{petromast12b}. Thus, in cases where the proton injection
is variable we find that the structure of the gamma-ray light curves
is complex due to the superposition of the intrinsic periodicity of the hadronic plasma
and of the variability pattern of the `external' source of proton injection.

\section{Summary}
We showed that the injection of high-energy protons,  e.g. $>10^{17}$~eV, 
with luminosities $\gtrsim 10^{52}$~erg/s in a region
which is part of a GRB-like flow with bulk Lorentz factor $\lesssim 500$
can lead to an abrupt energy transfer from protons to photons and neutrinos, which may carry
approximately $10\%-40\%$ of the injected proton luminosity.
We also found that the proton injection luminosity that is required for triggering the efficient cooling of protons
is higher,  e.g. reaching $10^{54}$~erg/s, for proton distributions extending to lower energies, e.g. $\gmx \sim 10^7$.
In this case, our model has the testable prediction of a contemporaneous bright burst  in $\gamma$-rays and high-energy neutrinos.
If the neutrino non-detection  from {\sl Fermi} bright GRBs with
IceCube will be established (e.g. \citealt{abbasi12, heetal12,liuwang13}), in the context of our model, it will mean
that the conditions in sources where proton acceleration to UHE is not possible, should 
be such as not to drive the system to supercriticality, e.g. $\Lpinj \ll 10^{54}$~erg/s.

We showed that in our framework the gamma-ray photon spectrum is self-consistently determined. 
Its shape is sensitive on how deep in the supercritical regime the system is driven. When 
the proton luminosity is marginally above the critical value, the photon spectrum 
peaks at $\sim 0.1$~GeV, whereas for higher proton luminosities it is modified
due to Comptonization and appears more as a Band spectrum. 
Not only different values of the proton luminosity but also small changes in one of the other parameters, and in particular
of the bulk Lorentz factor, may lead to photon spectra ranging between the two shapes described above.

Summarizing, we showed that supercriticalities are a generic feature of hadronic models, as
they manifest themselves for a wide range of parameters; from those relevant to AGN high-energy emission
to those relevant to GRBs. In the context of the latter, such supercriticalities 
can lead naturally to bursts of gamma-rays that share several properties with the typical GRB prompt emission, such as
the observed photon luminosity, the Band-like spectra and the time duration. These
offer also a unique way of transferring energy from protons to photons and neutrinos in a very efficient way,
and of overcoming the usual low-efficiency problem of hadronic models.

\section*{Acknowledgements}
Support for this work was provided by NASA 
through Einstein Postdoctoral 
Fellowship grant number PF 140113 awarded by the Chandra X-ray 
Center, which is operated by the Smithsonian Astrophysical Observatory
for NASA under contract NAS8-03060.
DG acknowledges support from the Fermi 6 cycle grant number 61122.

\bibliographystyle{mn2e} 
\bibliography{grbhadro} 
\appendix
\section[]{Derivation of critical proton compactness}
Here we derive the expression for the critical proton compactness ($\lpcr$) 
in the case where the gamma-rays that are spontaneously absorbed are the result of 
proton synchrotron radiation. We derive the expression for 
a power-law proton distribution with slope $p_{\rm p}=2$. A
similar analysis can be followed for different power-law indices.

The peak of the proton synchrotron spectrum in $\nu F_{\nu}$ units appears
at  
\eqb
\emx=b\mel c^2 \gamma_{\rm M}^2 \chi^{-1},
\label{emax}
\eqe
where $b=B/\Bcr$, $\Bcr=4.4\times10^{13}$~G,
$\chi=\mpr/\mel$ and $\gamma_{\rm M}=\min(\gmx, \gc)$. Here $\gc$ is the Lorentz factor
of protons that cool within the dynamical timescale $\rb/c$ and it is given by eq.~(\ref{gc}).
The maximum Lorentz factor of protons is $\gmx=\min(\gsat, \gh)$, where
$\gsat$ and $\gh$ are given by eqs.~(\ref{saturation}) and (\ref{hillas}), respectively.
\begin{itemize}
 \item The condition $\gh \le \gsat$ is translated to 
 \eqb
 \Gamma \gtrsim 140 \ \dt_{-1}^{-1/7} \eta_0^{1/7} \epsilon_{\rm B, -1}^{3/14} L_{\rm k, 52}^{3/14}
 \label{cond1}
 \eqe
In this regime, we find that $\gc > \gmx=\gh$ if 
\eqb
\Gamma \gtrsim 180 \dt_{-1}^{-1/7} \epsilon_{\rm B, -1}^{3/14} L_{\rm k, 52}^{3/14}
\label{cond2}
\eqe
where we used eqs.~(\ref{gc}) and (\ref{hillas}). Both conditions are identical apart from a numerical factor $1.3\eta_0^{-1/7}$.
Thus, as long as condition (\ref{cond1}) is satisfied we can assume that synchrotron proton cooling is not significant and
set $\gamma_{\rm M}= \gmx=\gh$.
\item The condition $\gh \ge \gsat$ is satisfied as long as 
\eqb
\Gamma \lesssim 140 \ \dt_{-1}^{-1/7} \eta_0^{1/7} \epsilon_{\rm B, -1}^{3/14} L_{\rm k, 52}^{3/14}.
\label{cond3}
\eqe
Using eqs.~(\ref{gc}) and (\ref{saturation}) we find that $\gc > \gsat$ only if
\eqb
\Gamma \gtrsim 190 \eta_0^{-1/7} \dt_{-1}^{-1/7} \epsilon_{\rm B, -1}^{-3/14} L_{\rm k, 52}^{3/14}.
\label{cond4}
\eqe
Inspection of conditions (\ref{cond3}) and (\ref{cond4}) reveals that in the regime where $\gmx=\gsat$ the high-energy
part of the proton distribution is affected by synchrotron losses, at least for most parameter values. 
For this,  we set $\gamma_{\rm M}=\gc$ as long as condition (\ref{cond3}) is satisfied.

\end{itemize}

The peak luminosity of the proton synchrotron spectrum in the comoving frame is then given by 
\eqb
L_{\rm p,syn} \approx L_0 N_{\rm p} B^2 \gamma_{\rm M},
\label{Lsyn}
\eqe
where
\eqb
L_0 \approx \frac{e^{9/2}}{\sqrt{6}\pi^{3/2} \mpr^{5/2} c^{7/2}}.
\eqe
$N_{\rm p}$ is the total number of protons which is related to the proton injection rate $Q_{\rm p}$ as
\eqb
N_{\rm p} & \approx & V t_{\rm esc} \int_1^{\gamma_{\rm M}}{\rm d}\gamma Q_{\rm p}(\gamma),
\label{Np}
\eqe
where  $V\simeq \pi \rb^3$, $t_{\rm esc}=\rb/c$ and $Q_{\rm p}=Q_0 \gamma^{-p_{\rm p}}$. The above expression is exact
only if $\gamma_{\rm M}=\gmx$. However, we find that is still a good approximation for $p_p > 1$ and $\gc \simeq (0.1-1) \gmx$.

Using eq.~(\ref{Np}) the proton injection compactness defined by eq.~(\ref{lp}) 
can be written in terms of $N_{\rm p}$ as
\eqb
\lp = \frac{\sth N_{\rm p}}{4\pi \rb^2} \ln(\gamma_{\rm M}).
\label{lpNp}
\eqe
Assuming that $ L_{\rm p, syn} \simeq L_{\gamma}$, where $L_{\gamma}$ is the integrated $\gamma$-ray luminosity, 
and using eqs.~(\ref{Lsyn}) and (\ref{lpNp}), the $\gamma$-ray compactness 
is written as
\eqb
\lph = \frac{L_0 \rb B^2 \gamma_{\rm M}}{\mel c^3 \ln(\gamma_{\rm M})}\lp .
\label{lph}
\eqe
As shown by \cite{petromast11}, the critical $\gamma$-ray compactness ($\lcr$) 
is a function of the $\gamma-$ray photon's energy (see eq.~(34) therein) having a minimum
at 
the energy (in $\mel c^2$ units)
\eqb
\xstar = 32^{2/9} \left(\frac{2}{b}\right)^{1/3}
\eqe
and increasing  as $\propto x^{1/2}$ for $x> x_\star$. The minimum value of $\lcr$
is found to be $\ell_{\star}=\left( 2^{7} b  \right)^{1/3}$. For $x\ge \xstar$, the critical $\gamma$-ray compactness 
may be written as
\eqb
\lcr = \ell_{\star}\left( \frac{x}{\xstar}\right)^{1/2}.
\label{lcr2}
\eqe
In general, the peak of the proton synchrotron spectrum ($x_{\max}$) appears at $x_{\max} \ge \xstar$. 
The transition to supercriticality occurs if $\lph \ge \lcr(x_{\max})$.
By combining eqs.~(\ref{B}), (\ref{emax}), (\ref{lph}), and (\ref{lcr2}) we find equivalently that
\eqb
 \lpcr  =   \frac{(2 \times 10^{-5})\Gamma_2^2}{ \epsilon_{\rm B,-1}^{1/2} L_{\rm k, 52}^{1/2}} \left\{ 
\begin{array} {ll}
22 + \ln \left( \frac{\epsilon_{\rm B, -1}^{1/2} L_{\rm  k, 52}^{1/2}}{\Gamma_2^2}\right), &  \gamma_{\rm M}=\gh \\
18 + \ln \left(\frac{\Gamma_2^5 \dt_{-1}} {\epsilon_{\rm B, -1} L_{\rm  k, 52}}\right), & \gamma_{\rm M}=\gc
\end{array}
\right.
\eqe

\section[]{Peak energy - Thomson depth relation}
The complete form of the Kompaneets equation \citep{kompaneets57} is 
\eqb
\frac{\partial \nph }{{\partial \tau}} = \tth\frac{\partial}{\partial x} \left[x^2\left(\nph + \frac{\nph^2}{x^2} + \Theta x^2 \frac{\partial \nph}{\partial x} \right) \right]
\eqe
where $\tau=ct/\rb$, $x=\epsilon / \mel c^2$, $\Theta=k T_e /\mel c^2$ and $\nph$ is the differential photon number density which is related to the photon occupation number 
$\mathcal{N}$ as $\mathcal{N} = (h c)^3 \nph(x) / 8 \pi x^2 (\mel c^2)^3$. 
In most astrophysically related cases $\mathcal{N} < 1$ and the induced emission term ($\propto \nph^2$) in the Kompaneets equation may be safely neglected.
Here we consider cases where $x \lesssim 1$ and $\Theta=0$, i.e. the Kompaneets equation describes the 
`recoil effect', where a photon loses energy through multiple scatterings with cold electrons. 

By adding a source and an escape term, the Kompaneets equation now reads
\eqb
\frac{\partial \nph }{{\partial \tau}} + \frac{\nph}{\tg(x,\tau)} = \tth(\tau)\frac{\partial}{\partial x}\left(x^2\nph(x)\right) + Q(x,\tau),
\label{kinetic}
\eqe
where $\tau$ is the time in $\rb/c$ units and $\tg$ is an approximate photon escape timescale given by
\eqb
\tg = 1 + \frac{\tth(\tau)}{3}f(x)
\eqe
and $f(x)$ is given by eq.~(21b) in \cite{lightmanzd87}. For our purposes, however, it is sufficient to use 
$f(x)=1$. Finally, the source term appearing in eq.~(\ref{kinetic}) is given by
\eqb
Q(x,\tau)=Q_0 x^{-s}H(1-\xmx)H(\xmx-x)H(x-\xmn) H(\tau).
\eqe
We note that the Thomson depth is, in principal, a time-dependent quantity. Using knowledge gained from the numerical study of the problem, 
according to which $\tth$ scales as $\tanh(\lambda \tau)$ with $\lambda > 1$ in the supercritical regime, we can consider $\tth$ to be constant.

Equation (\ref{kinetic}) is solved using the method of characteristics that transforms 
a partial differential equation (DE) into an ordinary DE along the characteristic curves or surfaces in two- or three- dimensional problems, respectively.
In our case, the characteristic curve is given by
\eqb
\frac{1}{x}-\frac{1}{x_0} = \tth(\tau-\tau_0),
\label{char}
\eqe
same as in the case of synchrotron or/and inverse Compton (in the Thomoson regime) cooling of relativistic electrons (see e.g. \citealt{kardashev62}).
First, we find the solution to the homogeneous equation by setting $Q=0$. Along the characteristic curve of eq.~(\ref{char}), the PDE
now reads
\eqb
\frac{1}{\nph}\frac{d\nph}{d\tau} = 2\tth x - \frac{1}{\tg},
\eqe
with the solution
\eqb
n_{\gamma, \rm H}(x,\tau) = \nph(x_0, \tau_0) \left(\frac{x_0}{x}\right)^2 \exp(-(\tau-\tau_0)/\tg).
\eqe
where the subscript $H$ stands for `homogeneous'. 
The solution to the equation including the source term is then
given by
\eqb
\nph(x,\tau) =\int_{-\infty}^{\tau}{\rm d}\tau_0  Q(x_0, \tau_0)\left(\frac{x_0}{x}\right)^2 e^{-(\tau-\tau_0)/\tg},
\eqe
where $x_0$ is a function of $\tau_0$ for fixed $x, \tau$ -- see eq.~(\ref{char}). 
The above integral is simplified by changing the integration variable from $\tau_0$ to $x_0$
using eq.~(\ref{char}):
\eqb
\nph(x,\tau) = \frac{Q_0 \tth}{x^2}e^{-a/x}\int_{x}^{x_{\rm M}}{\rm d}x_0 x_0^{-s} e^{a/x_0},
\label{sol}
\eqe
where $a^{-1}=\tth \tg$ and $x_{\rm M} = \min(\xmx, x_{\star})$ and $x_{\star} = (1/x-\tau \tg)^{-1}$.
The above integral is easily  calculated for a power-law injection with $s=2$ and it results in
\eqb
\nph(x, \tau) = \frac{Q_0 \tth}{a x^2} \left(1- e^{a(x_{\rm M}^{-1}-x^{-1})} \right),
\eqe
where 
\eqb
x_{\rm M} = \left\{ \begin{array}{cc}
                   \xmx, & x > \frac{\xmx}{1+\xmx \tth \tau} \\ \\
                   x_{\star}, & x \le \frac{\xmx}{1+\xmx \tth \tau}
                   \end{array}
	      \right.
\eqe
The peak of $x^2 \nph(x,\tau)$ appears at $x_{\rm p} = \xmx/(1+\xmx \tth \tau)$. Since photons escape
from the source in an average time $\tau \simeq \tg$, the peak energy is given
by
\eqb
x_{\rm p} \simeq \frac{\xmx}{\xmx+\tth \tg} \simeq \frac{3}{\tth^2},
\eqe
where the last equation holds for  $\tth \gg 1$.
\end{document}